\documentclass[graybox,envcountchap]{svmult}

\usepackage{aas_macros,amsmath,amssymb,cite,esint,graphicx,hyperref,mathptmx,tensor,tikz}
\usepackage[misc]{ifsym}
\hypersetup{colorlinks=true,urlcolor=blue}

\newcommand{\ed}{\mathop{}\!\mathrm{d}}
\newcommand{\ab}[1]{\left|#1\right|}
\newcommand{\br}[1]{\left[#1\right]}
\newcommand{\pa}[1]{\left(#1\right)}
\newcommand{\av}[1]{\left\langle#1\right\rangle}
\newcommand{\ind}[1]{\indices{#1}}
\DeclareMathOperator{\sign}{sign}

\begin{document}

\title{A Beginner's Guide to Black Hole Imaging and Associated Tests of General Relativity}
\titlerunning{A Beginner's Guide to Black Hole Imaging}

\author{Alexandru Lupsasca, Daniel R. Mayerson, Bart Ripperda, and Seppe Staelens}
\institute{Alexandru Lupsasca \at Department of Physics and Astronomy, Vanderbilt University, Nashville, TN 37212, USA,\\
\email{alexandru.v.lupsasca@vanderbilt.edu}
\and 
Daniel R. Mayerson (\Letter) \at Institute for Theoretical Physics, KU Leuven, Celestijnenlaan 200D, B-3001 Leuven, Belgium,\\
\email{daniel.mayerson@kuleuven.be}
\and 
Bart Ripperda \at Canadian Institute for Theoretical Astrophysics, Toronto, ON M5S 3H8, Canada,\\
Department of Physics, University of Toronto, Toronto, ON M5S 1A7, Canada,\\
D.~A.~Dunlap Department of Astronomy, University of Toronto, Toronto, ON M5S 3H4, Canada,\\
Perimeter Institute for Theoretical Physics, Waterloo, ON N2L 2Y5, Canada,\\
Flatiron Institute, Center for Computational Astrophysics, New York, NY 10010, USA,\\
\email{ripperda@cita.utoronto.ca}
\and
Seppe Staelens \at Department of Applied Mathematics and Theoretical Physics, Centre for Mathematical Sciences, University of Cambridge, Wilberforce Road, Cambridge CB3 0WA, United Kingdom,\\
\email{ss3033@cam.ac.uk}}

\maketitle

\abstract{Following the 2019 release by the Event Horizon Telescope Collaboration of the first pictures of a supermassive black hole, there has been an explosion of interest in black hole images, their theoretical interpretation, and their potential use in tests of general relativity.
The literature on the subject has now become so vast that an introductory guide for newcomers would appear welcome.
Here, we aim to provide an accessible entry point to this growing field, with a particular focus on the black hole ``photon ring'': the bright, narrow ring of light that dominates images of a black hole and belongs to the black hole itself, rather than to its surrounding plasma.
Far from an exhaustive review, this beginner's guide offers a pedagogical review of the key basic concepts and a brief summary of some results at the research frontier.}

\numberwithin{equation}{section}

\tableofcontents
\clearpage

\section{Introduction}

Despite their ubiquity, black holes are very mysterious.
As the strongest sources of gravity, they have long represented the ultimate frontier of gravitational physics---an arena where Einstein's theory of General Relativity (GR) is pushed to its limit, and most likely beyond.
According to GR and its celebrated ``no-hair'' theorems, the spacetime geometry of a stationary black hole is completely described by the Kerr metric, which---remarkably---is entirely specified by only two parameters: the black hole mass and its spin.
Hence, from this classical perspective, black holes are the simplest macroscopic objects.\footnote{On the other hand, the Bekenstein-Hawking area law, which states that the black hole entropy is proportional to the area of the event horizon (rather than its enclosed volume), suggests that they are the most information-dense, and hence complex, quantum systems in the universe.
This apparent contradiction suggests that the description of black holes in GR is incomplete; its resolution is a driving goal in the pursuit of quantum gravity, in which one may hope for some experimental input.}
In principle, this striking simplicity, coupled with their extreme gravitational fields, makes black holes a prime target for tests of GR.
For a long time, however, most electromagnetic observations of black holes have been limited---by necessity---to resolving only the light emitted relatively far away from the black hole environment.
As a result, it has proved difficult, if not impossible, to use electromagnetism to probe this gravitational frontier with very much precision.

The Event Horizon Telescope (EHT) collaboration has dramatically shifted this paradigm with its first observations via Very-Long-Baseline Interferometry (VLBI) of the supermassive black holes at the centers of the galaxy M87 \cite{EHT2019I} and of our galaxy (Sgr~A$^*$) 
\cite{EHT2022I}.
For the first time, EHT observations achieved the frequency (230\,GHz) required to resolve the horizon-scale emission from the vicinity of these sources.
Excitingly, this observed light likely includes photons whose near-horizon trajectories deviated from their Newtonian paths (straight lines) in the most dramatic way, describing nearly bound spherical orbits around the black hole.

Naturally, the advent of high-frequency VLBI has been met with a resurgence of interest into how such electromagnetic observations could be used to carry out precision tests of GR.
This question is becoming all the more pressing as improved observations are expected to be delivered in the near future, from both the ground---with the next-generation EHT (ngEHT) \cite{Johnson2023,Ayzenberg2023}---and from space \cite{Gurvits2022}.
However, to the best of our knowledge, the current literature seems to lack a basic introductory text on this topic---one that explains the basic theory behind black hole imaging and provides a first look at the most promising ideas for testing GR with VLBI observations, while remaining mindful of their inherent (and often-overlooked) potential pitfalls.
Our goal in this Chapter is precisely to provide such a pedagogical tool.
As a result, this text is aimed primarily at students (or to novices in the area), and does not purport to give a comprehensive review of the field.
Rather, this Chapter is our personal tour through the subject---one necessarily biased by our own experiences, interests, and expertise.

\bigskip

Tests of GR via electromagnetic observations of black holes pose a number of challenges.
Perhaps the greatest difficulty arises from the need to disentangle the effects produced by the black hole's gravity from those caused by specific features of the particular source emitting the photons that we eventually detect.
At first glance, one would expect that significant uncertainties in the detailed properties of the astrophysical source would always be present and dwarf any observable deviation from GR, rendering the very idea of a test of Einstein's theory in this context moot.

Fortunately, even when the source of emission is imperfectly known (as discussed in Section~\ref{sec:Plasma}), there exist clever ways to nevertheless extract robust geometrical data from black hole imaging observations.
In our opinion, the most promising idea is to measure the \emph{photon ring}: the bright ring in a black hole image that surrounds the central dark depression (or ``shadow'') caused by the event horizon. This photon ring actually consists of an infinite sequence of subrings, with the $n^\text{th}$ subring formed by photons that executed $\sim n$ half-orbits around the black hole before escaping its gravitational pull.
Observable features of the photon ring have been argued to be robust across many different astrophysical models of the emission, establishing the photon ring as the part of a black hole image that belongs to (and is determined by) \emph{the black hole geometry itself}, rather than the surrounding plasma.
For this reason, a large part of our discussion will be centered around properties of the photon ring, which we regard as the most promising avenue for extracting geometrical data from black hole imaging---and thus testing GR.

An analogy can help to illustrate how the photon ring disentangles the geometric, gravitational effects from the emission source details.\footnote{We thank Andrew Strominger for this analogy \cite{Podcast}.}
Consider a clothing store with a changing room that is completely covered in mirrors.
When staring at the mirror directly in front, the \emph{direct image} one sees will merely show the clothes that one is wearing; and conversely, one needs to know the details of one's outfit to be able to predict that image.
On the other hand, the same mirror will also display a sequence of higher-order mirrored images arising from multiply reflected light.
These carry information about the structure of the \emph{room} itself: by comparing the relative sizes and positions of successive images, one can in principle reconstruct the \emph{geometry} of the room and the arrangements of the mirrored walls completely. 

It is in precisely the same way that the direct ($n=0$) image reveals the astrophysics of the source emitting around a black hole (corresponding in the previous analogy to the ``clothes'' that one is wearing), while the structure of the photon ring---composed of the higher-order ($n\geq 1$) lensed images---reveals information about the underlying black hole spacetime geometry (the ``structure of the room'').

\bigskip

It is useful to reflect on the exact goal that one may have in mind when extracting geometrical data from an observation in order to ``test GR''.
Standard reasoning in Effective Field Theory (EFT) \emph{a priori} predicts that corrections to GR should only become important at curvature scales that are many orders of magnitude larger than those that can be achieved outside of a supermassive black hole's event horizon (see Section~\ref{sec:BeyondEinstein}).
As a result, our theoretical prior for GR ought to be extremely strong.
Thus, from an EFT perspective, it is very difficult to conceive how any black hole imaging observation could lead to conclusive evidence that GR is ``wrong''.\footnote{An exception to this reasoning arises when a supermassive black hole is in fact a hitherto-unknown, compact, horizonless object.
However, if such objects are to give rise to noticeably distinct and observable features, they should presumably be very \emph{large} compared to their horizon scales---otherwise, their ``structure'' will still be confined within the light ring and thus largely unobservable via imaging.
Intuitively, it is then doubtful that we would be able to observe horizon-scale corrections to black hole geometry, such as those proposed in the fuzzball paradigm or by reflective ``mirrors'' at the horizon; see Section~\ref{sec:BeyondEinstein} and references therein for further discussion.} 

This does not imply that it is useless to carefully analyze observations of the geometry around imaged black holes.
Indeed, we have very few other observations of the ``strong gravity'' produced by such large gravitational potentials and curvatures; GR has not yet been confirmed in this regime, which remains largely unexplored.
Even if we do not \emph{a priori} expect any deviations from GR, it is imperative to confirm our best theory of gravity in the most extreme regimes that we can probe.
However, it does (very likely) mean that we will only be able to perform meaningful \emph{null tests, or consistency tests}, in which we can show GR to be consistent with the observed data up to a certain precision level; it is less clear if or how black hole imaging can lead to \emph{discriminatory tests}, where a (physically viable) alternative to GR is shown to be preferred over GR by the observed data.
We discuss this further in Section~\ref{sec:Testing} (especially in Sections~\ref{sec:NullTests} and \ref{sec:testexamples}).

Nevertheless, even a mostly qualitative confirmation of the presence of orbiting light would constitute a new observation and a valuable consistency test of GR.
Of course, it should be mentioned that a mission to resolve this orbiting light would also provide other measurements of great astrophysical interest---such as a more accurate estimate of the black hole spin---as well as insights into phenomena such as black hole electromagnetic energy extraction or jet launching.

Finally, in thinking about the detailed theoretical properties of the photon ring, and by following the long narrow thread connecting them to actual observables, we can hope to learn new ways of thinking about black holes.
This seems inherently valuable and will undoubtedly shed new light on old problems in black hole physics.

\bigskip

As mentioned above, this Chapter does \emph{not} aim to provide an exhaustive review of all the possible topics related to tests of GR with black hole imaging.
Instead, it is intended to guide ``beginners'' in this field towards understanding the most salient features of black hole imaging.
We especially want to emphasize the difficulties in extracting geometrical data from black hole images and the associated potential pitfalls; we feel that these are often overlooked, but are nevertheless key to understanding the GR tests that black hole imaging can deliver.
Finally, we sketch a tour through some interesting ideas and results that we find particularly promising in this context.
Another review of GR tests with black hole imaging is \cite{Psaltis2019}; it can be regarded as complementary to ours since we focus on different aspects of this topic.

The rest of this Chapter is structured as follows.
In Section~\ref{sec:Basics}, we review some of the basic physics underlying black hole imaging.
We devote special attention to the theory of null geodesics orbiting black holes (Section~\ref{sec:Theory}), and explain how their existence is responsible for the emergence of the photon ring (Section \ref{sec:PhotonRing}).
We also cover the basic mechanics behind VLBI observations (Section~\ref{sec:Interferometry}), and emphasize the uncertainties in the astrophysical source of emission (Section~\ref{sec:Plasma}).

Then, in Section~\ref{sec:Observables}, we provide an overview of some recent, state-of-the-art proposals for interferometric observations of certain image features (mostly connected to the photon ring) that could in principle be used to test GR.
Lastly, in Section~\ref{sec:Testing}, we discuss how these features could actually lead to tests of GR, dwelling further on the difference (introduced above) between consistency versus discriminatory tests.
Finally, we conclude with a brief outlook on the future of black hole imaging with VLBI observations. Throughout the text, we use units in which $c=G=1$.

\section{Pedagogical review of key concepts}
\label{sec:Basics}

This Section is devoted to a pedagogical introduction to the basic principles of black hole imaging.
We start with the theory of null geodesics reaching an observer screen (Section~\ref{sec:Theory}), and describe how nearly bound geodesics give rise to photon rings and their properties (Section~\ref{sec:PhotonRing}).
We then review the observational technique that is used to form black hole images (Section~\ref{sec:Interferometry}) and describe the unknowns in these images due to the inherent uncertainties about the emitting plasma (Section~\ref{sec:Plasma}).

\subsection{Black hole images and the Kerr critical curve}
\label{sec:Theory}

The ``image'' of a black hole is created from light emitted by radiating matter in its vicinity, which is strongly influenced by its gravitational pull.
In this Section, we review a general framework in which to describe such an image and its key features.

\subsubsection*{The observer screen and null geodesics in the Kerr geometry}

A general stationary and axisymmetric metric in vacuum general relativity has a line element of the form
\begin{align}
    ds^2=-e^{2\nu}\ed t^2+e^{2\psi}\pa{\ed\phi-\omega\ed t}^2+e^{2\lambda}\ed r^2+e^{2\mu}\ed\theta^2\,,
\end{align}
where the functions $\nu$, $\psi$, $\omega$, $\lambda$, and $\mu$ depend only on the poloidal coordinates $r$ and $\theta$, and not on the toroidal coordinates $t$ and $\phi$.
A test particle with four-momentum $p_\mu=g_{\mu\nu}\dot{x}^\nu$ (dots denote derivatives with respect to an affine parameter) following a geodesic path will have associated conserved quantities $E\equiv-p_t$ and $L\equiv p_\phi$, which respectively correspond to the particle's energy and azimuthal angular momentum about the spin axis of the black hole.
Additionally, a third constant of motion is provided by the normalization condition $p_\mu p^\mu=-\mu^2$ for a test particle of mass $\mu$.

\begin{figure}[tb]
    \centering
    \begin{tikzpicture}[thick,scale=0.7]
        \draw (-2, 0) arc (-180:180:2);
        \draw (8, -4) -- (11, -3) -- (11, -0.5) -- (8,-1.5) -- (8,-4);
        \draw (8, -2.75) -- (11, -1.75) node[anchor = west]{\large $\alpha$};
        \draw (9.5, -3.5) -- (9.5, -1) node[anchor = south]{\large $\beta$};
        \draw (-2, 0) arc (-180:0:2 and 0.7);
        \draw (0, -0.7) -- (-0.3,-0.4); 
        \draw (0, -0.7) -- (-0.3,-1); 
        \draw (0, -2.8) -- (0, -2);
        \draw (0,2) -- (0, 2.8);
        \draw[red, thick] (-1, 0.5) -- (8, -1.84);
        \draw[red, thick, dashed] (8, -1.84) -- (9, -2.1);
        \draw[red, thick] (-1, 0.5) arc (-130:-230:4.6 and 1.5);
        \draw[red, thick] (-1, 2.8) -- (2, 3.58);
        \node at (9, -2.1) {\textbullet};
        \node at (2, 3.58) {\textbullet};
        \node at (2.4, 3.7) {$D$};
        \draw (9.5, -2.25) -- (9,-2.1); 
        \draw (9., -2.1) arc (140:45:0.7 and 0.7);
        \node at (9.2, -2.6) {$R$};
        \node at (10.1, -1.5) {$\phi_R$};
    \end{tikzpicture}
    \caption[Geometry of an observer screen aligned with a black hole.]{An observer screen with impact parameters $(\alpha,\beta)$ located far away from the black hole, at spatial coordinates $(r_{\rm o},\theta_{\rm o},\phi_{\rm o})$.
    The $\beta$ axis is defined to be the black hole spin axis projected onto the plane perpendicular to the line of sight to the center of the black hole, and the $\alpha$ axis is defined to be perpendicular to it.
    A test particle that follows a geodesic (red) originating from the distant point $D$ strikes the observer screen after being deflected by the black hole.
    The coordinates $(R,\phi_R)$ are polar coordinates on the observer screen.}
    \label{fig:ObserverScreen}
\end{figure}
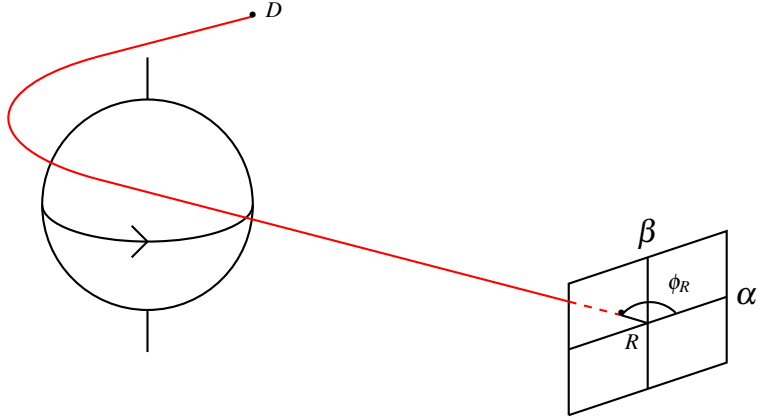

In this spacetime, let us imagine an observer with a screen whose center points towards the center of symmetry of the asymptotic gravitational field; see Figure~\ref{fig:ObserverScreen}.
This observer sets up a frame of reference represented by the orthonormal tetrad
\begin{align*}
    e_{(t)}&=e^{-\nu}\br{\frac{\partial}{\partial t}+\omega\frac{\partial}{\partial\phi}}\,,&
    e_{(\phi)}&=e^{-\psi}\frac{\partial}{\partial\phi}\,,\\
    e_{(r)}&=e^{-\lambda}\frac{\partial}{\partial r}\,,&
    e_{(\theta)}&=e^{-\mu}\frac{\partial}{\partial\theta}\,.
\end{align*}
In this frame, the locally measured energy of the test particle is equal to
\begin{align}
    p^{(t)}\equiv-e_{(t)}^\mu p_\mu
    =e^{-\nu}\pa{E-\omega L},
\end{align}
while the observed momentum in the directions tangent to the observer screen is
\begin{align}
    p^{(\phi)}=e^{-\psi}L\,,\qquad
    p^{(\theta)}= e^{-\mu}p_\theta\,.
\end{align}

On the observer screen, we can use Cartesian coordinates $\alpha$ and $\beta$ to represent the impact parameters in the directions perpendicular and parallel to the projected axis of symmetry of the spacetime, respectively.
For an observer who is located at large radius $r_{\rm o}$ and inclination $\theta_{\rm o}$ relative to the symmetry axis of the spacetime, the impact parameters of the test particle are given by the asymptotic formulae \cite{Bardeen1973,Gralla2018}
\begin{align}
    \label{eq:GeneralCoordinates}
    \alpha=\lim_{r_{\rm o}\to\infty}-r_{\rm o}\left.\frac{p^{(\phi)}}{p^{(t)}}\right|_{\theta=\theta_{\rm o}}\,,\qquad
    \beta=\lim_{r_{\rm o}\to\infty}r_{\rm o}\left.\frac{p^{(\theta)}}{p^{(t)}}\right|_{\theta=\theta_{\rm o}}\,.
\end{align}
Alternatively, one could also opt for dimensionless polar coordinates $(R,\cos{\phi_R})$ on the observer screen, defined using the (dimensionless) angles $\alpha/r_{\rm o}$ and $\beta/r_{\rm o}$:
\begin{align}
    R=\frac{\sqrt{\alpha^2+\beta^2}}{r_{\rm o}}\,,\qquad
    \cos{\phi_R}=\frac{\alpha}{Rr_{\rm o}}\,.
\end{align}

Spacetimes in which geodesics admit a fourth (independent) conserved quantity---besides $E$, $L$, and $\mu$ --allow for their geodesic motion to be completely integrated.
We focus on the Kerr geometry, for which Carter \cite{Carter1968} was the first to demonstrate this property.
In Boyer-Lindquist coordinates $(t,r,\theta,\phi)$, the Kerr line element is
\begin{gather}
    ds^2=-\frac{\Delta}{\Sigma}\pa{\ed t-a\sin^2{\theta}\ed\phi}^2+\frac{\Sigma}{\Delta}\ed r^2+\Sigma\ed\theta^2+\frac{\sin^2{\theta}}{\Sigma}\br{\pa{r^2+a^2}\ed\phi-a\ed t}^2\,,\notag
\end{gather}
where $M$ is the mass parameter, $a \in [-1,1]$ is the dimensionless spin parameter and
\begin{gather}
    \Delta=r^2-2Mr+a^2,\qquad
    \Sigma=r^2+a^2\cos^2{\theta}.
\end{gather}
The fourth independent constant of geodesic motion in Kerr---known as the \emph{Carter constant}---is given by \cite{Carter1968}
\begin{align}
    k&=-\Delta p_r^2-r^2\mu^2+\frac{1}{\Delta}\br{-\pa{r^2+a^2}E+aL}^2\\
    &=p_\theta^2+a^2\mu^2\cos^2{\theta}+\frac{1}{\sin^2{\theta}}\br{L-aE\sin^2{\theta}}^2\,.
\end{align}
These two expressions are equivalent when using $p_\mu p^\mu=-\mu^2$.
It is often more convenient to use the constant $Q\equiv k-(L-aE)^2$ instead.

The existence of this fourth independent constant of the motion can be traced back to the existence in the Kerr geometry of an irreducible, rank-two, symmetric Killing tensor \cite{Walker1970}.
Many other (black hole) spacetimes possess such a higher-rank Killing symmetry, and thus a fourth constant of geodesic motion.
If instead the metric only admits a conformal Killing tensor, then the fourth constant of geodesic motion exists only for null geodesics; two examples of this are the Rasheed-Larsen black hole \cite{Rasheed1995,Larsen2000,Keeler2012}, or the related so-called almost-BPS black hole \cite{Bah2021,Staelens2022}.

We now focus on Kerr light rays, that is, Kerr null geodesics with $\mu=0$.
One can directly solve for the components of the geodesic four-momentum $p^\mu=\dot{x}^\mu$ by imposing conservation of the four quantities $E$, $L$, $Q$, and $\mu$: in terms of specific (energy-rescaled) conserved quantities $\lambda\equiv L/E$ and $\eta\equiv Q/E^2$, this yields \cite{Carter1968,GrallaLupsasca2020b}
\begin{align}
    \label{eq:RadialTrajectory}
    \frac{\Sigma}{E}\frac{dr}{d\sigma}&=\pm_r\sqrt{\mathcal{R}(r)}\,,\\
    \label{eq:PolarTrajectory}
    \frac{\Sigma}{E}\frac{d\theta}{d\sigma}&=\pm_\theta\sqrt{\Theta(\theta)}\,,\\
    \label{eq:AzimuthalTrajectory}
    \frac{\Sigma}{E}\frac{d\phi}{d\sigma}&=\frac{a}{\Delta}\pa{r^2+a^2-a\lambda}+\frac{\lambda}{\sin^2{\theta}}-a\,,\\
    \label{eq:TemporalTrajectory}
    \frac{\Sigma}{E}\frac{dt}{d\sigma}&=\frac{r^2+a^2}{\Delta}\pa{r^2+a^2-a\lambda}+a\pa{\lambda-a\sin^2{\theta}}\,,
\end{align}
where we have introduced the angular potential
\begin{align}
    \label{eq:AngularPotential}
    \Theta(\theta)\equiv\frac{p_\theta^2}{E^2}
    =\eta+a^2\cos^2{\theta}-\lambda^2\cot^2{\theta}\,,
\end{align}
as well as the radial potential 
\begin{align}
    \label{eq:RadialPotential}
    \mathcal{R}(r)\equiv\Delta^2\frac{p_r^2}{E^2}
    =\pa{r^2+a^2-a\lambda}^2-\Delta\br{\eta+\pa{\lambda-a}^2}\,,
\end{align}
while $\pm_r$ and $\pm_\theta$ denote the signs of $p^r$ and $p^\theta$, respectively.
From Eqs.~\eqref{eq:PolarTrajectory} and \eqref{eq:AngularPotential}, we see that when a geodesic passes through the equatorial plane, its (squared) instantaneous momentum perpendicular to the equatorial plane $p_\theta^2$ equals $\Theta(\pi/2)=\eta$.
Therefore, all geodesics that cross the equator must have $\eta\geq0$.

Using Eqs.~\eqref{eq:RadialTrajectory}--\eqref{eq:TemporalTrajectory}, the expressions \eqref{eq:GeneralCoordinates} for the impact parameters on the screen of a Kerr observer become \cite{Bardeen1973,Chandrasekhar1992}\footnote{The original expression for $\beta$ in Eq.~(42b) of \cite{Bardeen1973} is missing the square root and a factor $a^2$.}
\begin{align}
    \label{eq:KerrAlpha}
    \alpha&=-\frac{\lambda}{\sin{\theta_{\rm o}}}\,,\\
    \label{eq:KerrBeta}
    \beta&=\sign\pa{p_{\rm o}^\theta}\sqrt{\eta+a^2\cos^2{\theta_{\rm o}}-\lambda^2\cot^2{\theta_{\rm o}}}\,,
\end{align}
where $p_{\rm o}^\theta$ denotes the photon polar momentum at the observer, or equivalently,
\begin{align}
    R&=r_{\rm o}^{-1}\sqrt{a^2\cos^2{\theta_{\rm o}}+\eta+\lambda^2}\,,\\
    \label{eq:KerrPhi}
    \phi_R&=\sign\pa{p_{\rm o}^\theta}\arccos\pa{-\frac{\lambda}{Rr_{\rm o}\sin{\theta_{\rm o}}}}\,. 
\end{align}
For Kerr---as for any spacetime with separable null geodesics---there is therefore a one-to-one correspondence between the conserved quantities $(\lambda,\eta)$ of a photon and its impact parameters $(\alpha,\beta)$ on the observer screen.

\subsubsection*{Parallel transport of linear polarization}

The linear polarization of a photon (relevant for Section~\ref{sec:PhotonRingPolarization}) is described by a four-vector $f^\mu$ perpendicular to the direction of light propagation (so $f\cdot p=0$) that is parallel-transported along the null geodesic (so $p^\mu\nabla_\mu f^\nu=0$).
The direction of linear polarization can be conveniently packaged in the complex Penrose--Walker constant $\kappa$, which is also a conserved quantity along null geodesics \cite{Connors1977,Connors1980}:
\begin{align} 
    \label{eq:PenroseWalker}
    \kappa&=\kappa_1+i\kappa_2
    =\pa{\mathcal{A}-i\mathcal{B}}\pa{r-ia\cos{\theta}},\\
    \mathcal{A}&=\pa{p^t f^r-p^r f^t}+a\sin^2{\theta}\pa{p^r f^\phi-p^\phi f^r},\\
    \mathcal{B}&=\br{\pa{r^2+a^2}\pa{p^\phi f^\theta-p^\theta f^\phi}-a\pa{p^t f^\theta-p^\theta f^t}}\sin{\theta}.
\end{align}
The conservation of $\kappa$ follows from the existence of a conformal rank-two antisymmetric Killing-Yano tensor in Kerr, so that this construction can in principle also be generalized to any spacetime admitting such a symmetry \cite{Chandrasekhar1992}.
Applications of this technology to the parallel transport of linear polarization in Kerr include, e.g., \cite{Li2009,Dexter2016,Lupsasca2018,Himwich2020,Narayan2021,Gelles2021}.

\subsubsection*{Polar geodesic motion}

The polar behavior of null geodesics can be completely determined from the form of the angular potential \eqref{eq:AngularPotential}, which by \eqref{eq:PolarTrajectory} must always be positive along a trajectory: $\Theta(\theta)\geq0$.
This implies that no geodesic can reach the poles at $\theta=0$ and $\theta=\pi$ unless $\lambda=0$.\footnote{We will not consider the special case $\lambda=0$ further here (see, e.g., \cite{Kapec2020} for more discussion), as it corresponds to a measure-zero set of points on the observer screen.}

For $\lambda\neq0$, the polar motion is restricted to lie between turning points $\theta_\pm$ where $\Theta(\theta_\pm)=0$.
In terms of $u=\cos^2{\theta}$, we can rewrite the angular potential \eqref{eq:AngularPotential} as
\begin{align}
    \label{eq:PolarPotential}
    \Theta(u)=\frac{a^2}{1-u}\pa{u_+-u}\pa{u-u_-},
\end{align}
where we defined the roots
\begin{align}
    \label{eq:PolarRoots}
    u_\pm=\triangle_\theta\pm\sqrt{\triangle_\theta^2+\frac{\eta}{a^2}}\,,\qquad
    \triangle_\theta=\frac{1}{2}\pa{1-\frac{\eta+\lambda^2}{a^2}}\,.
\end{align}
There are two possible types of polar trajectories \cite{Kapec2020,GrallaLupsasca2020b}:
\begin{itemize}
    \item Ordinary geodesics with $\eta>0$.
    These have two physical roots
    \begin{align}
        \label{eq:TurningPoints}
        \theta_1=\arccos\pa{\sqrt{u_+}}\,,\qquad
        \theta_4=\arccos\pa{-\sqrt{u_+}}\,,
    \end{align}
    with $0<\theta_1<\pi/2<\theta_4<\pi$ and the potential positive between them.
    Hence the photon oscillates between $\theta_1$ and $\theta_4$, crossing the equatorial plane each time.
    \item Vortical geodesics with $\eta<0$.
    These have two additional roots
    \begin{align}
        \theta_2=\arccos\pa{\sqrt{u_-}}\,,\qquad
        \theta_3=\arccos\pa{-\sqrt{u_-}}\,,
    \end{align}
    with $0<\theta_1<\theta_2<\pi/2<\theta_3<\theta_4<\pi$ and the potential positive over the ranges $(\theta_1,\theta_2)$ and $(\theta_3,\theta_4)$, which allows for motion that lies purely within the northern and southern hemisphere, respectively.
\end{itemize}

\subsubsection*{Bound photon orbits and the Kerr critical curve}

Light rays that are traced backwards into the geometry from the observer screen can fall into the black hole or be deflected by its gravitational pull but eventually escape to infinity.
The boundary delineating these two regions consists of null geodesics that asymptote to bound photon orbits that circumnavigate the black hole forever; their apparent position on the observer screen forms a closed curve known as the \emph{critical curve}, whose interior (resp. exterior) corresponds to the region of photon capture (resp. photon escape).
Bardeen \cite{Bardeen1973} completely characterized the bound photon orbits in the Kerr spacetime and analytically described the critical curve on the observer screen $(\alpha,\beta)$.\footnote{Remarkably, for the simpler case of a Schwarzschild black hole, the critical curve was first derived by mathematician David Hilbert in 1917 \cite{Hilbert1917}, only mere months after Schwarzschild's discovery of his eponymous solution in 1916, which itself directly followed the 1915 publication of Einstein's field equations.
The subject then seemed to be largely forgotten until Darwin's 1959 work \cite{Darwin1959}.}
We now briefly review its derivation.

Bound photon orbits are null geodesics that remain at a constant radius $r$.\footnote{This is not obvious a priori, as one could imagine bound orbits that oscillate between two different radii.
However, a careful analysis of the possible turning points of $\mathcal{R}$ (see, e.g., Section IV of \cite{GrallaLupsasca2020b}) leads to the conclusion that Kerr bound photon orbits can only occur at fixed $r$; in fact, this can be viewed as the defining characteristic of the Boyer-Lindquist radial coordinate in the Kerr geometry.}
This corresponds to the twin conditions
\begin{align}
    \label{eq:BoundOrbits}
    \mathcal{R}(r)=\mathcal{R}'(r)=0\,,
\end{align}
which respectively ensure that $\dot{r}=0$ and $\ddot{r}=0$.
These two conditions fix\footnote{The second set of solutions to \eqref{eq:BoundOrbits} is $\lambda=\frac{r^2+a^2}{a}$ and $\eta=-\frac{r^4}{a^2}$, which is unphysical because the angular potential---and hence $p_\theta^2$ --is negative for this solution: $\Theta(\theta)=-\pa{\frac{\Sigma}{a\sin{\theta}}}^2<0$.} the specific conserved quantities $\lambda$ and $\eta$ in terms of the orbital radius $r$:
\begin{align}
    \label{eq:CriticalParameters}
    \lambda=\frac{M\pa{r^2-a^2}-r\Delta}{a(r-M)}\,,\qquad
    \eta=\frac{4r^2\Delta}{(r-M)^2}-\pa{\lambda-a}^2\,. 
\end{align}
Inserting these expressions into Eq.~\eqref{eq:PolarRoots} yields the roots
\begin{align}
    u_\pm(r)=\frac{r}{a^2(r-M)^2}\br{-r^3+3M^2r-2a^2M\pm2\sqrt{M\Delta\pa{2r^3-3Mr^2+a^2M}}}\,.\notag
\end{align}
Bound orbits cannot be vortical because $u_-(r)<0$ everywhere outside the horizon, resulting in unphysical angles $\theta_{2,3}\notin\br{0,\pi}$.
Therefore, bound orbits are necessarily ordinary and oscillate in the range $\br{\theta_1,\theta_4}$, which is bounded by the turning points \eqref{eq:TurningPoints} and always includes the equatorial plane.
As mentioned above, crossing the equator requires $\eta\geq0$, so this is a necessary condition for these orbits to be allowed.
As a result, the radii of the innermost and outermost bound orbits can be found by solving \eqref{eq:CriticalParameters} for $\eta=0$.
Defining $r'\equiv r-2M$, this amounts to solving
\begin{align}
    r'^3-3M^2r'+2M^3-4a^2M=0\,.
\end{align}
This is a depressed cubic equation with three distinct solutions
\begin{align}
    r_k=2M\pa{1+\cos\br{\frac23\arccos\pa{\pm\frac{a}{M}}-\frac{2\pi k}{3}}}\,,\qquad
    k\in\{0,1,2\}\,.
\end{align}
The two solutions outside of the event horizon are
\begin{align}
    r_\pm^\gamma=2M\pa{1+\cos\br{\frac{2}{3}\arccos\pa{\pm\frac{a}{M}}}}\,,
\end{align}
with $r_-^\gamma$ corresponding to the innermost orbital radius, and $r_+^\gamma$ to the outermost one.

\begin{figure}[ht]
    \centering
    \includegraphics[width=0.5\textwidth]{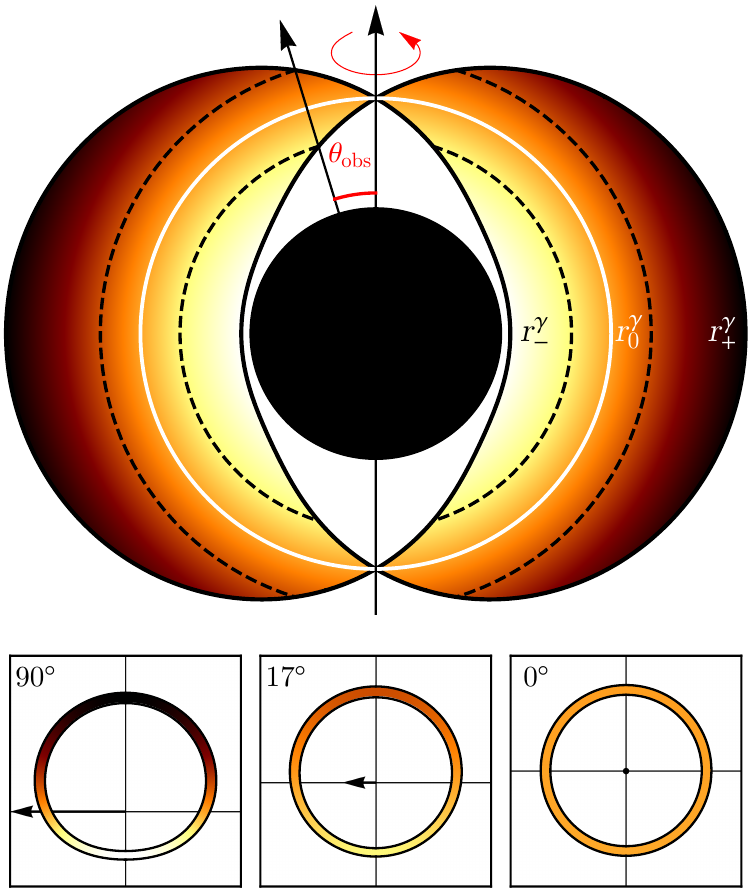}
    \caption{\emph{Top:} The photon shell of a Kerr black hole with spin $a/M=0.94$.
    The boundaries $r_\pm^\gamma$ are shown, as well as the zero-angular-momentum orbit at $r_0^\gamma=r_3$.
    An observer at $\theta_{\rm o}=17^\circ$ only observes the part of the photon ring that corresponds to the intersection of his/her line of sight with the photon sphere.
    \emph{Bottom:} The photon ring as viewed by observers at different inclinations.
    The color matches that of the radii that contribute to the image.
    Only an observer in the equatorial plane observes the photon ring associated with the entire photon shell.
    The face-on observer at $\theta_{\rm o}=0^\circ$ only observes photons associated with the bound orbit at $r_{\rm o}$.
    The arrow denotes the projection of the spin axis on the plane perpendicular to the line of sight.
    This is Figure~2 from \cite{JohnsonLupsasca2020}, reproduced under the terms of the Creative Commons Attribution 4.0 International License.}
    \label{fig:PhotonShell}
\end{figure}

Bound photon orbits therefore exist in the range $r_-^\gamma\le r\le r_+^\gamma$, and they orbit between polar angles $\theta_-\le\theta\le\theta_+$ with
\begin{align}
    \label{eq:OrbitalTurningPoints}
    \theta_\pm(r)=\arccos\pa{\mp\sqrt{u_+(r)}}\in[0,\pi],
\end{align}
and sweep out every azimuthal angle $\phi\in[0,2\pi)$.
These bounds define the region of spacetime containing bound null geodesics, which is referred to as the \emph{photon shell}; see Figure~\ref{fig:PhotonShell} for a visualization.

An analytic formula for the \emph{critical curve} on the observer screen can be obtained in parametric form (with the orbital radius $r$ as parameter) by inserting the critical impact parameters \eqref{eq:CriticalParameters} into the expressions \eqref{eq:KerrAlpha}--\eqref{eq:KerrBeta} for the screen coordinates $(\alpha,\beta)$.
For an observer in the equatorial plane ($\theta_{\rm o}=\pi/2$), the full range $r_-^\gamma\le r\le r_+^\gamma$ of bound orbits contributes to the critical curve; at lower inclinations, the critical curve is parameterized by the subset of this interval over which \eqref{eq:KerrBeta} remains real.
This discussion implies (rather counter-intuitively) that different \emph{angles} around the critical curve on the observer screen correspond not to different angles $\phi$ in the spacetime geometry, but rather to different orbital \emph{radii} in the photon shell: this is a striking manifestation of the warped nature of the Kerr spacetime!

\begin{figure}[ht]
    \centering
    \includegraphics[width=0.49\textwidth]{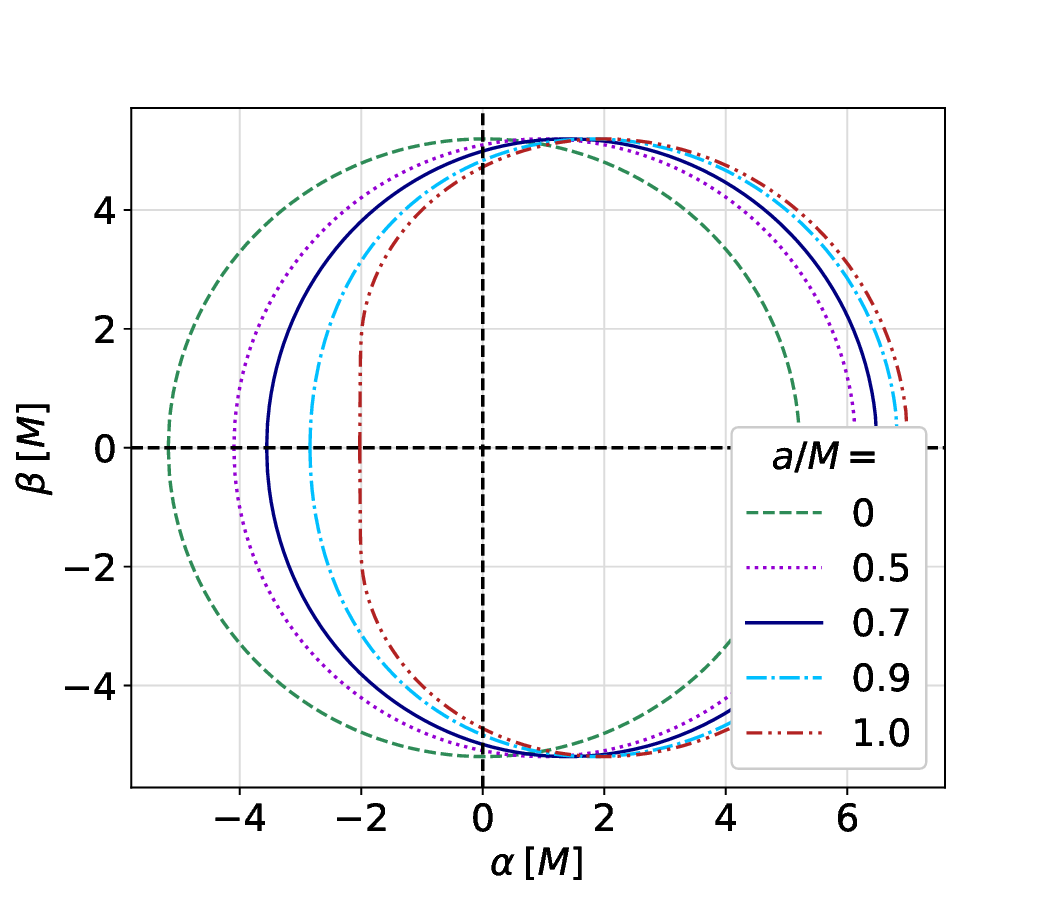}
    \includegraphics[width=0.49\textwidth]{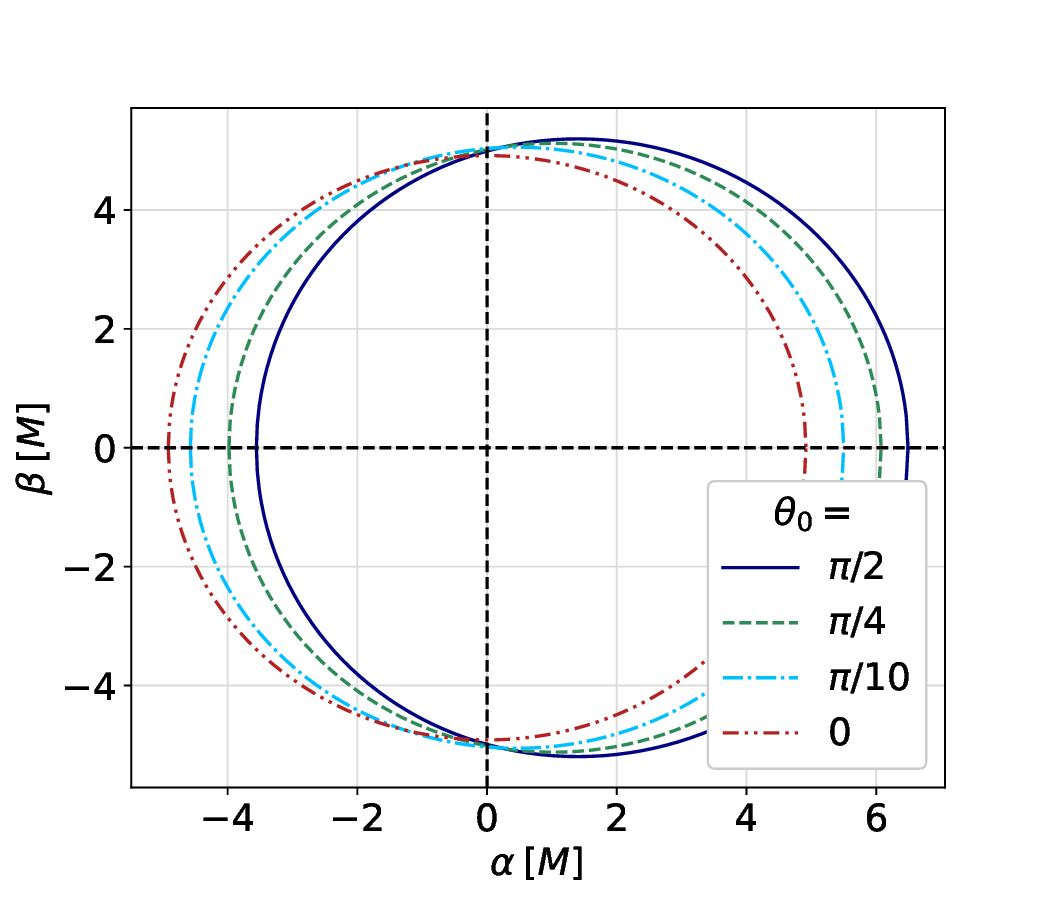}
    \caption{\emph{Left:} The Kerr critical curve for different values of the black hole spin parameter at an observer inclination $\theta_{\rm o}=90^\circ$.
    \emph{Right:} The Kerr critical curve for different observer inclinations $\theta_{\rm o}$ in the case of a black hole with spin $a/M=0.7$.
    The solid curves in both panels correspond to the same critical curve.}
    \label{fig:KerrCriticalCurve}
\end{figure}

The Kerr critical curves corresponding to different black hole spins and observer inclinations are illustrated in Figure~\ref{fig:KerrCriticalCurve}.
The critical curve becomes a circle either when the observer lies on the spin axis (due to axisymmetry), or when the black hole is not rotating (due to spherical symmetry).
The observability of the critical curve (or its lack thereof) will be discussed in  Section~\ref{sec:Shadow} below.

\subsection{Nearly bound orbits, photon rings and critical exponents}
\label{sec:PhotonRing}

The poloidal components \eqref{eq:RadialTrajectory}--\eqref{eq:PolarTrajectory} of the geodesic equation imply that
\begin{align}
    \pm_r\frac{1}{\sqrt{\mathcal{R}(r)}}\frac{dr}{d\sigma}=\frac{E}{\Sigma}
    =\pm_\theta\frac{1}{\sqrt{\Theta(\theta)}}\frac{d\theta}{d\sigma}\,.
\end{align}
Integrating this equation along a geodesic from affine parameter $\sigma=\sigma_i$ to $\sigma=\sigma_f$ --corresponding to coordinates $(t_i,r_i,\theta_i,\phi_i)$ and $(t_f,r_f, \theta_f,\phi_f)$, respectively---yields
\begin{align}
    \label{eq:MinoTime}
    \fint_{r_i}^{r_f}\frac{\ed r}{\pm_r\sqrt{\mathcal{R}(r)}}=\fint_{\theta_i}^{\theta_f}\frac{\ed\theta}{\pm_\theta\sqrt{\Theta(\theta)}}\,,
\end{align}
where the slash through each integral indicates that it is to be evaluated along the photon trajectory, with the signs $\pm_r$ and $\pm_\theta$ chosen to ensure that each integral grows monotonically along the path.
These integrals can be reduced to Legendre normal form, that is, they are expressible in terms of three standard elliptic integrals \cite{GrallaLupsasca2020b} (see also, e.g., \cite{Rauch1994,Vazquez2004,Li2005}).

We now focus on bound photon orbits with polar turning points \eqref{eq:OrbitalTurningPoints} and let $G_\theta$ denote the polar integral over one oscillation from $\theta_-$ to $\theta_+$ (a half period),
\begin{align}
    G_\theta\equiv\int_{\theta_+}^{\theta_+}\frac{\ed\theta}{\sqrt{\Theta(\theta)}}
    =2\int_{\pi/2}^{\theta_+}\frac{\ed\theta}{\sqrt{\Theta(\theta)}}\,,
\end{align}
where the last line follows from the fact that $\theta_+=\pi-\theta_-$, with the polar oscillation symmetric about the equatorial plane.
Using Eq.~\eqref{eq:PolarPotential}, we can rewrite this as
\begin{align}
    \label{eq:HalfOrbitIntegral}
    G_\theta=\frac{1}{\sqrt{a^2}}\int_0^{u_+}\frac{\ed u}{\sqrt{u\pa{u_+-u}\pa{u-u_-}}}
    =\frac{2}{\sqrt{-a^2u_-}}K\!\pa{\frac{u_+}{u_-}}\,,
\end{align}
where we introduced the complete elliptic integral of the first kind,
\begin{align}
    K(m)\equiv\int_0^1\frac{\ed t}{\sqrt{\pa{1-t^2}\pa{1-m\, t^2}}}\,.
\end{align}
By construction, $G_\theta>0$ represents the polar integral along a complete half-orbit from $\theta_\pm$ to $\theta_\mp$.

\subsubsection*{Nearly bound photons}

\begin{figure}
    \centering
    \includegraphics[width=0.8\textwidth]{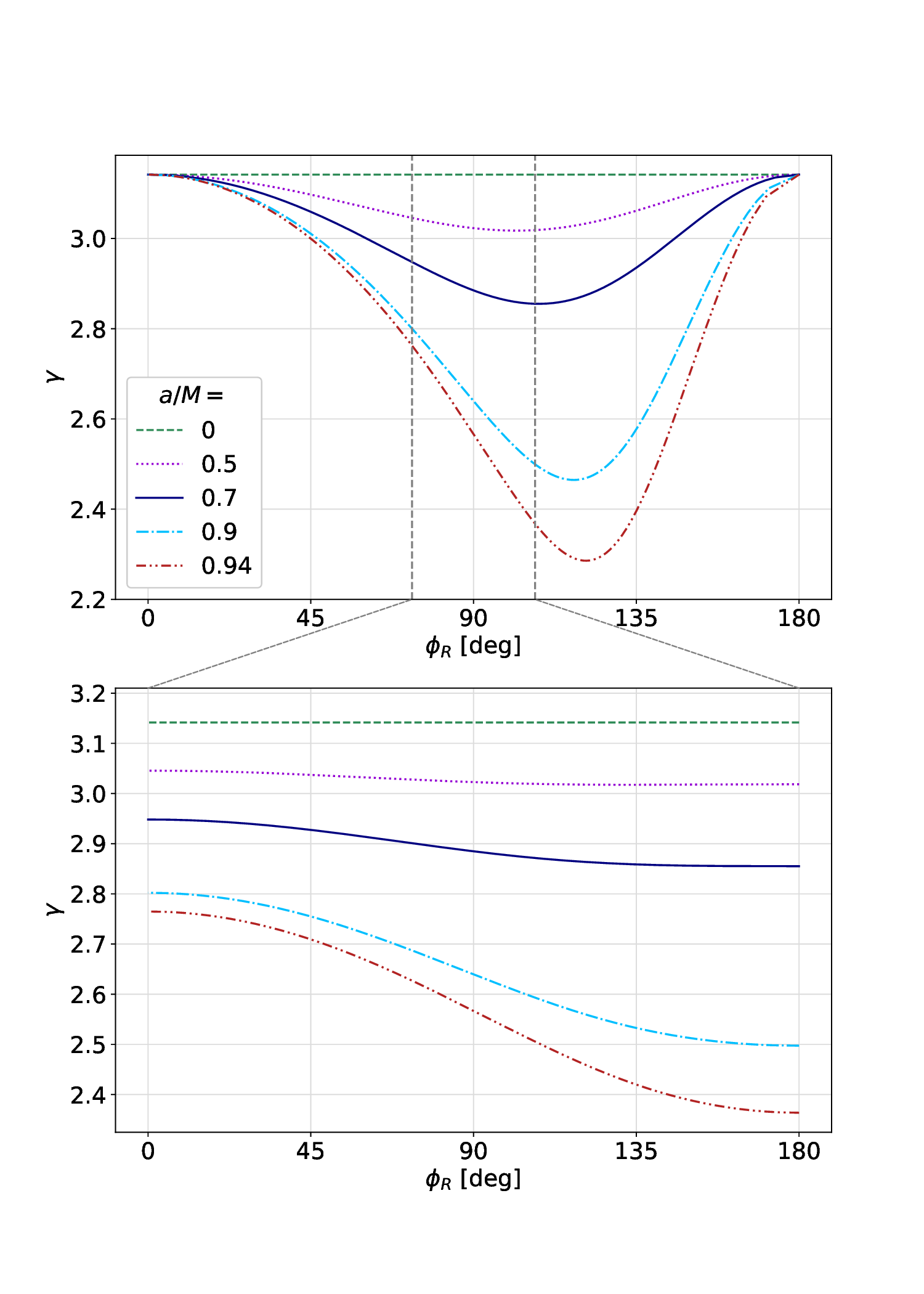}
    \caption{The Lyapunov exponent $\gamma$ as function of the polar angle $\phi_R$ around the observer screen, for different values of the spin parameter $a/M$ in the Kerr metric.
    The function $\gamma(\phi_R)$ accessible to an observer at inclination $\theta_{\rm o}=17^\circ$ \emph{(bottom)} is a subset of the full range accessible to an equatorial observer \emph{(top)}.}
    \label{fig:KerrLyapunovExponent}.
\end{figure}

Consider now a photon that is bound at orbital radius $r=\tilde{r}$ within the photon shell $r^\gamma_-\le\tilde{r}\le r^\gamma_+$, with conserved quantities \eqref{eq:CriticalParameters}.
If we give it a slight displacement $\delta r_0$, such that its initial radius is
\begin{align}
    r_i=\tilde{r}+\delta r_0\,,\qquad
    0<\delta r_0\ll\tilde{r}\,,
\end{align}
then the photon becomes unbound.
As long as the photon remains near its orbital shell, its radial potential can be approximated as
\begin{align}
    \label{eq:NearlyBound}
    \mathcal{R}(r)\approx\frac{1}{2}\mathcal{R}''(\tilde{r})\pa{\delta r_0}^2\,,
\end{align}
since $\mathcal{R}(\tilde{r})=\mathcal{R}'(\tilde{r})=0$ by Eq.~\eqref{eq:BoundOrbits} applied to the bound orbit at radius $\tilde{r}$.
For simplicity, we suppose that the photon starts in the equatorial plane at $\theta_i=\pi/2$.
After completing $n$ polar half-orbits between $\theta_\pm$ and $\theta_\mp$, the photon has moved to a radius $r_f=\tilde{r}+\delta r_n$ that is determined by the poloidal geodesic equation \eqref{eq:MinoTime},
\begin{align}
    \int_{\tilde{r}+\delta r_0}^{\tilde{r}+\delta r_n}\frac{\ed r}{\sqrt{\mathcal{R}(r)}}=nG_\theta(\tilde{r})\,.
\end{align}
Using \eqref{eq:NearlyBound}, the radial integral can be approximated as
\begin{align}
    \int_{\tilde{r}+\delta r_0}^{\tilde{r}+\delta r_n}\frac{\ed r}{\sqrt{\mathcal{R}(r)}}&\approx\int_{\tilde{r}+\delta r_0}^{\tilde{r}+\delta r_n}\frac{\ed r}{\sqrt{\frac{1}{2}\mathcal{R}''(\tilde{r})(r-\tilde{r})^2}}
    =\sqrt{\frac{2}{\mathcal{R}''(\tilde{r})}}\ln\pa{\frac{\delta r_n}{\delta r_0}}\,. 
\end{align}
Hence, the radial deviation $\delta r=r-\tilde{r}$ of the photon from its nearby bound orbit grows exponentially (as long as $\delta r$ is small).
This exponential instability, which is characteristic of unstable fixed points, may be quantified by the \emph{Lyapunov exponent}
\begin{align}
    \gamma\equiv\sqrt{\frac{\mathcal{R}''(\tilde{r})}{2}}G_\theta(\tilde{r})\,,
\end{align}
which controls the exponential rate of radial deviation,
\begin{align}
    \delta r_n=e^{\gamma n}\delta r_0\,.
\end{align}
Using the expressions \eqref{eq:RadialPotential}, \eqref{eq:CriticalParameters}, and \eqref{eq:HalfOrbitIntegral}, we obtain an explicit expression for the \textbf{Lyapunov exponent} of Kerr photon orbits,
\begin{align}
    \boxed{
    \gamma(r)=\frac{4}{\sqrt{-a^2u_-}}\sqrt{r^2-\frac{Mr\Delta}{(r-M)^2}}K\!\pa{\frac{u_+}{u_-}}
    \,.}
\end{align}
This Lyapunov exponent is a function of the black hole mass $M$, its spin $a$, and the photon orbital radius $r$.
At every such radius, we can invert $\phi_R(r)$ from Eq.~\eqref{eq:KerrPhi} to obtain the orbital radius $r(\phi_R)$ corresponding to polar angle $\phi_R$ around the critical curve.
Thus, we may also regard the Lyapunov exponent as a function $\gamma(\phi_R)$ of the polar angle around the observer screen.
This is illustrated in Figure~\ref{fig:KerrLyapunovExponent}.

As described in our earlier discussion, off-equatorial observers only probe a subset of the bound orbits in the photon shell $r_-^\gamma\le r\le r_+^\gamma$.
As a result, it follows that the function $\gamma(\phi_R)$ computed for an equatorial observer contains all the information encoded in $\gamma(\phi_R)$ computed for off-equatorial observers (but the converse is not true).
The angles $\phi_R^\pm$ between which an equatorial observer probes the radii observable from inclination $\theta_{\rm o}$ obey the approximate relation \cite{Staelens2023}
\begin{align}
    \phi_R^\pm\approx\pa{\frac{\pi}{2}\pm\theta_{\rm o}}\mod\pi\,.
\end{align}
Figure~\ref{fig:KerrLyapunovExponent} illustrates this for an observer at inclination $\theta_{\rm o}=17^\circ$, for whom $\phi_R^-\approx73^\circ$ and $\phi_R^+\approx107^\circ$.

\subsubsection*{The photon ring}

The existence of nearly bound photon orbits is responsible for the appearance of a striking and universal feature in black hole images: the \emph{photon ring} \cite{Beckwith2005,Gralla2019,JohnsonLupsasca2020}.
This bright, narrow ring arises from light rays that are strongly bent by the gravitational pull of the black hole and orbit through its surrounding emission multiple times---collecting additional photons in each pass through the emission---before reaching the observer and producing a brightness enhancement close to the critical curve.

Consider a point with coordinates $(R_c,\phi_R)$ on the critical curve.
A light ray that is aimed outside the critical curve at a small perpendicular distance $\delta R_n>0$  from this point will typically cross the equatorial plane a number of times of order \cite{JohnsonLupsasca2020}
\begin{align}
    n\approx-\frac{1}{\gamma(\phi_R)}\ln\pa{\frac{\delta R_n}{R_c}}\,.
\end{align}
It follows that for this light ray to complete an additional half-orbit around the black hole, it must be aimed exponentially closer to the critical curve, at a distance
\begin{align}
    \label{eq:ExponentialDemagnification}
    \delta R_{n+1}\approx e^{-\gamma(\phi_R)n}\delta R_n\,.
\end{align}

The photon ring then consists of an infinite series of subrings, such that the $n^\text{th}$ subring contains photons that completed a fixed number $n$ of equatorial crossings on their way from source to observer.
As described by Eq.~\eqref{eq:ExponentialDemagnification}, the Lyapunov exponent governs how each successive subring grows exponentially closer to the critical curve and moreover decreases exponentially in width.
For a Schwarzschild black hole, this phenomenon was first understood by Darwin \cite{Darwin1959} and numerically investigated by Luminet \cite{Luminet1979}.

The photon ring is an important and ubiquitous feature in black hole images.
The photon subrings' precise locations on the observer screen depend on the emission responsible for them, although the $n^\text{th}$ subring will always lie in a well-defined $n^\text{th}$ ``lensing band'': a region of the observer screen consisting of all the possible light rays that could achieve at least $n$ half-orbits around the black hole on their trajectory \cite{Paugnat2022,CardenasAvendano2023}.
If the black hole is surrounded by an optically thin equatorial disk emitting photons isotropically, then a geodesic that crosses the equatorial plane (and therefore also the accretion disk) $n$ times will contain $\sim n$ times more photons compared to a geodesic that only crosses the disk once.
Therefore, the intensity of rays lying in high-$n$ photon subrings is expected to be larger by a factor $\sim n$ as well \cite{Gralla2019,JohnsonLupsasca2020}.
We will discuss the interferometric signatures of the photon ring in Section~\ref{sec:InterferometricPhotonRing} and some more of its observable features in Sections~\ref{sec:PhotonRingPolarization}, \ref{sec:PhotonRingShape}, and \ref{sec:PhotonRingAutocorrelations}.

\subsubsection*{Critical exponents of the photon shell and ring}

We complete the discussion of (nearly) bound geodesics and the photon ring by describing the two critical exponents $\delta$ and $\tau$ that characterize the toroidal motion of bound photons.

Following \cite{GrallaLupsasca2020a} (see also \cite{Teo2003}), we can integrate the geodesic equations \eqref{eq:AzimuthalTrajectory} and \eqref{eq:TemporalTrajectory} for $\phi$ and $t$ along a geodesic bound at a fixed orbital radius $r$ to obtain:
\begin{align}
    \Delta\phi=a\pa{\frac{r+M}{r-M}}\int\frac{\ed\theta}{\pm_\theta\sqrt{\Theta(\theta)}}+\lambda\int\frac{\csc^2{\theta}}{\pm_\theta\sqrt{\Theta(\theta)}}\ed\theta\,.
\end{align}
Over a complete half-orbit ($\theta_\pm$ to $\theta_\mp$), this change in $\phi$ can be denoted by $\hat{\delta}$.
However, $\hat{\delta}$ is not a smooth function of the bound orbit radius $r$, as it has a $2\pi$ jump discontinuity at the pole-crossing bound orbit $r_{\rm o}$ for which $\lambda=0$ \cite{Teo2003,GrallaLupsasca2020a}.
It is therefore more natural to define a smooth function $\delta=\hat{\delta}+2\pi H(r-r_{\rm o})$, or more explicitly,
\begin{align}
    \delta=2a\pa{\frac{r+M}{r-M}}G_\theta+\lambda\int_{\theta_-}^{\theta_+}\frac{\csc^2{\theta}}{\sqrt{\Theta(\theta)}}\ed\theta+2\pi H(r-r_{\rm o})\,,
\end{align}
where we used the Heaviside step function $H(x)$, and $G_\theta$ defined in \eqref{eq:HalfOrbitIntegral}.
This expression defines the critical exponent $\delta$ for Kerr; by definition, it determines the $\phi$ motion of the bound orbit at radius $r$ (and nearly bound geodesics at radii $r+\delta r$).

In an entirely similar way (although without the complication involving the pole-crossing orbit $r_{\rm o}$), we can convert the $t$ integral coming from \eqref{eq:TemporalTrajectory} to an angular integral over $\theta$; the resulting integral over one complete half-orbit defines the critical exponent $\tau$:
\begin{align}
    \tau=2r^2\pa{\frac{r+3M}{r-M}}G_\theta+a^2\int_{\theta_-}^{\theta_+}\frac{\cos^2{\theta}}{\sqrt{\Theta(\theta)}}\ed\theta\,.
\end{align}
These expressions for $\delta$ and $\tau$ can be written in terms of elliptic integrals \cite{GrallaLupsasca2020a}.
The critical exponents $\gamma$, $\delta$, and $\tau$ control black hole lensing \cite{GrallaLupsasca2020a} and play a crucial role in the description of photon ring brightness autocorrelations \cite{Hadar2021}; see Section~\ref{sec:PhotonRingAutocorrelations}.

\subsubsection*{Extension to other spacetimes}

In this Section (and in Section~\ref{sec:Theory}), we have focused on the calculation of bound null orbits, the critical curve, and the critical exponents $\gamma$, $\delta$, and $\tau$ governing the (nearly) bound geodesics that make up the photon rings of Kerr black holes.
All of these calculations can be extended to other black hole metrics, as long as they admit a separable null geodesic equation (and thus a fourth, Carter-like conserved quantity);\footnote{Of course, the details of each step may change.
For example, one must check that the only bound orbits occur at fixed radius, i.e., that no radially oscillating bound orbits exist.
Additionally, for spacetimes that are not equatorially symmetric, the angular integral from the equator to $\theta_+$ will not be equal to the integral from the equator to $\theta_-$ (and additionally $\theta_+\neq\pi-\theta_-$ for such spacetimes); see \cite{Staelens2023} for some examples.}
see \cite{Staelens2022, Staelens2023, Wielgus2021,Ayzenberg2022,Omwoyo2023,Olmo2023} for some examples.

In spacetimes where the null geodesic equation is not separable, bound photon orbits will generically still exist, although it is perhaps not clear how a critical curve can be defined unambiguously (see \cite{Staelens2023} for an example).
Recently, a formalism for calculating Lyapunov exponents in spacetimes with non-separable geodesics was developed in \cite{Deich2023}, which exploits the definition of the Lyapunov exponent as the measure of instability along the trajectory of a bound orbit in geodesic phase space.

\subsection{Interferometric black hole imaging}
\label{sec:Interferometry}

The Event Horizon Telescope (EHT) images black holes using a technique known as Very-Long-Baseline Interferometry (VLBI).
That is, it is a radio interferometer composed of antennas located at the different telescope sites in the array.
Each of these antennas measures a local electric field as a function of time $t$, frequency $\nu$, and polarization $P$; we can then consider the complex cross-correlation between the measured electric field for each pair of antennas $(i,j)$, a quantity known as the interferometric visibility:
\begin{align}
    V_{ij}(t,\nu,P_1,P_2)=\av{E_i(t,\nu,P_1)E_j^*(t,\nu,P_2)}\,.
\end{align}
There are four possible such cross-correlations, corresponding to the four different combinations of polarization $(P_1,P_2)$.
We will focus here on the total (brightness) intensity of the image, which corresponds to one particular combination of these correlations \cite{Roberts1994} (see Section~\ref{sec:PhotonRingPolarization} for a discussion of the observed polarization, which corresponds to a different cross-correlation).

Every pair of antennas in the array thus samples the complex visibility $V$, which is related by the van Cittert-Zernike theorem \cite{vanCittert1934,Zernike1938} to the Fourier transform of the image brightness $I$ in the plane of the sky: if $\mathbf{x}=(x,y)$ are dimensionless Cartesian coordinates in the image plane (measured, for instance, in radians), then
\begin{align}
    V(\mathbf{u})=\int I(\mathbf{x})e^{-2\pi i\mathbf{u}\cdot\mathbf{x}}\ed^2\mathbf{x}\,,
\end{align}
where the dimensionless ``baseline'' $\mathbf{u}$ can be viewed as a vector projected onto the plane perpendicular to the line of sight to the source, and measured in units of the observation wavelength $\lambda=c/\nu$.
Each baseline $\mathbf{u}$ joining two antennas in the interferometric array samples a complex visibility $V(\mathbf{u})$, which corresponds to a single Fourier component of the sky image $I(\mathbf{x})$.
Technical and environmental limitations make the phase of the visibility difficult to measure; on the other hand, it is comparably easier to access the visibility amplitude $|V|$ (sometimes also called the ``visamp'').
More details on interferometry can be found in \cite{Roberts1994,EHT2019IV}, for instance.

\begin{figure}[ht]
    \centering
    \includegraphics[width=0.49\textwidth]{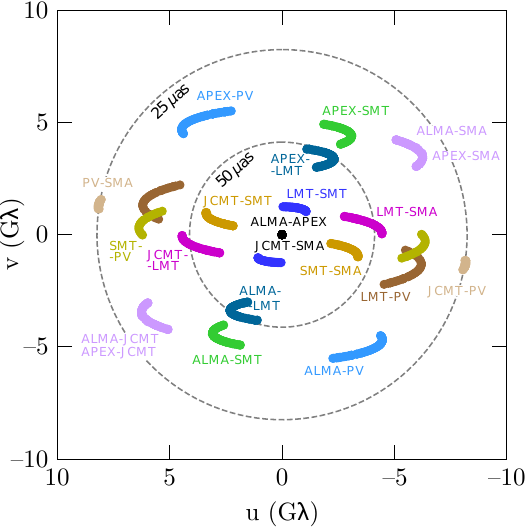}
    \includegraphics[width=0.49\textwidth]{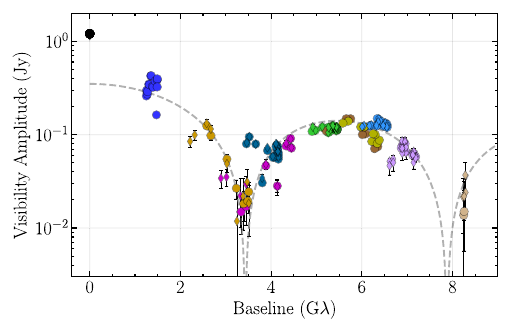}
    \caption{\emph{Left:} Plot of the visibility points in the baseline plane $\mathbf{u}=(u,v)$ sampled by the different pairs of telescopes in the EHT array in its 2017 observations of M87.
    The tracks are produced by the Earth's rotation, which helps fill out the baseline plane (this is known as ``Earth rotation aperture synthesis'').
    \emph{Right:} The corresponding measured visibility amplitudes (overlaid with a Fourier transform of a symmetric thin ring model, display as a dashed line). This is Figure~2 of \cite{EHT2019I}.}
    \label{fig:baselinecoverage}
\end{figure}

\begin{figure}[ht]
    \centering
    \includegraphics[width=\textwidth]{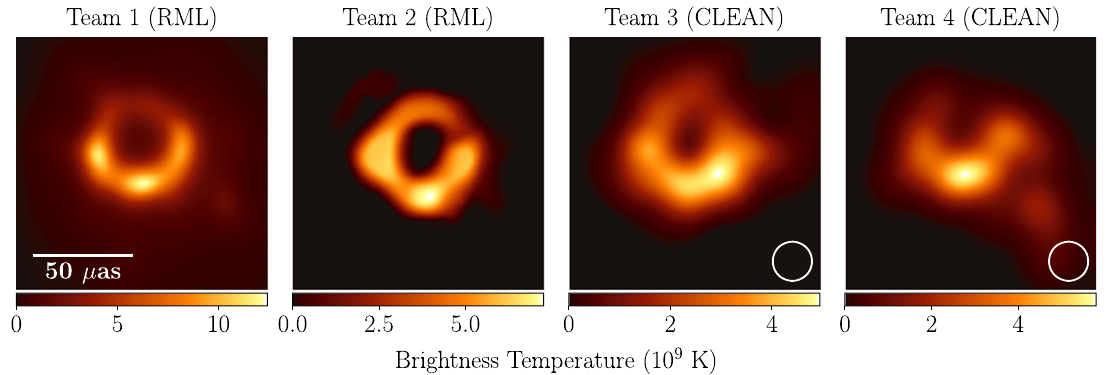}
    \caption{Four different images reconstructed by independent EHT imaging teams from the sparsely sampled interferometric data collected in the 2017 EHT observations of M87.
    Each image was produced by a different image reconstruction algorithm (named above its panel).
    This is Figure~4 of \cite{EHT2019IV}, where these different algorithms are described.}
    \label{fig:FirstImages}
\end{figure}

Each pair of antennas only measures the visibility $V$ at a given baseline $\mathbf{u}$ (which changes somewhat over time due to the rotation of the Earth); see Figure~\ref{fig:baselinecoverage}.
We clearly never have access to the entire visibility function $V(\mathbf{u})$: this means that, even in principle, we do not have enough information to completely reconstruct the image brightness $I(\mathbf{x})$ from its sparsely sampled Fourier transform.
Therefore, any ``image'' produced from EHT observations necessarily relies on assumptions and modeling.
This point is well-illustrated by the first four ``images'' produced independently by the four imaging teams within the EHT collaboration from the first observations of M87 (see Figure~\ref{fig:FirstImages}): the same sparse interferometric data is consistent with a wide range of reconstructed images \cite{EHT2019IV}.
Only the features that are consistent across all images (reconstructed via different algorithms) should be treated as robust; other features are likely artifacts of a particular reconstruction method.

\begin{figure}[ht]
    \centering
    \includegraphics[width=\textwidth]{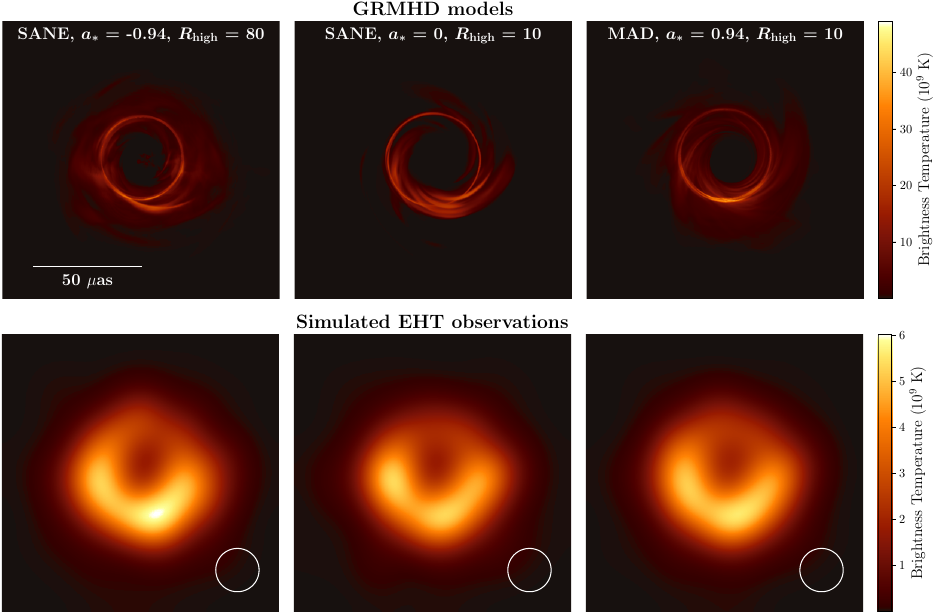}
    \caption{Three snapshots \emph{(top)} and their corresponding simulated EHT observations \emph{(bottom)} raytraced from GRMHD simulations with different prescriptions for the electron heating (which is modeled in post-processing and not from first principles).
    This is Figure~4 of \cite{EHT2019I}.}
    \label{fig:DifferentSourceModels}
\end{figure}

In addition to the image uncertainty due to sparse interferometric sampling, there is an additional uncertainty in the physics of the emitting plasma in the accretion flow \cite{EHT2019V,EHT2022V}: Figure~\ref{fig:DifferentSourceModels} shows images obtained from simulated EHT observations of three very different emission models for the plasma in the accretion disk.
We further discuss this physical uncertainty in Section~\ref{sec:Plasma} below.

It is also important to mention the timescales involved in these observations.
In particular, the EHT images of M87$^*$ shown in Figures~\ref{fig:FirstImages} and \ref{fig:DifferentSourceModels} can be understood as \emph{time-averaged} images.
The EHT observations range over a number of days; since the plasma around the black holes evolves in time, its emission changes during the observation period.
The timescale of the plasma dynamics is likely correlated with the black hole's mass and set by its light-crossing time; for M87$^*$, this is roughly $GM/c^3\approx 9$ hours, so the effect is still rather mild in the EHT's multi-day observations of M87$^*$ \cite{EHT2019I}.
On the other hand, the black hole at the center of our galaxy, Sgr~A$^*$, is about 1,500 times lighter than M87$^*$; its dynamical timescale is on the order of $GM/c^3\approx 21$ seconds.
Accordingly, the EHT observations of Sgr~A$^*$ do indeed show significant variability in the image over its measurement period \cite{EHT2022I,EHT2022IV}.
This variability makes it more challenging to time-average observational data and leads to larger uncertainties in the reconstructed image compared to M87$^*$.
At the same time, it could also be an asset for GR tests, as it produces autocorrelations (between fluctuations in the direct image and their light echoes) whose measurement would provide a sharp probe of the underlying geometry---see discussion in Section~\ref{sec:PhotonRingAutocorrelations}.

\bigskip

There are clearly many uncertainties in the interferometrically sampled image and the underlying plasma physics.
To extract information about the underlying geometry---that is, to access information that can potentially test GR---the relevant question is then: \emph{What features of a black hole ``image'' are universal, i.e., shared by all (or at least, most) interferometric image reconstruction methods as well as physical emission models?}
Only such universal features can reliably be used to extract robust geometrical data about the black hole geometry and thus lead to potential tests of gravitational physics.

Perhaps the most obvious universal feature of the produced black hole images (see Figures~\ref{fig:FirstImages} and \ref{fig:DifferentSourceModels}) is the existence of a bright ring-like structure surrounding a dark brightness depression (often called the ``shadow'', but see further below in Section~\ref{sec:Shadow}).
Such a bright ring reveals itself in the actual interferometric signal as a Bessel-function-like behavior of the visibility amplitude (see Figure~\ref{fig:baselinecoverage}).
The presence of a bright ring surrounding a central brightness depression is itself a nontrivial prediction of GR, which certainly indicates the existence of an ultracompact object at the center of the source \cite{EHT2019I}, and allows for an estimate of this object's mass \cite{EHT2019VI}.\footnote{This mass can typically be measured in other independent ways, for example by observing stellar motions around the object.
The EHT mass measurement can then confirm these measurements (as a rough test of GR), or alternatively assume them to constrain other parameters \cite{EHT2019VI}.}
The asymmetry of the image is also fairly robust and can be attributed to a nonzero spin of the object, although it is harder to make a quantitative estimate of this angular momentum from the image \cite{EHT2019V}.

Of course, the mass and angular momentum of the central object are only its most basic properties.
To really test GR, we need to be able to extract more detailed properties and features of the underlying geometry
We will discuss this prospect further in Section~\ref{sec:Observables}; in particular, we will argue that the most promising method for extracting such geometric properties is from precise interferometric features of the bright ring in the image, that is, from the photon ring.

\subsection{Plasma physics uncertainties in simulations of the emission}
\label{sec:Plasma}

In the Section above, we discussed the interferometric observations that eventually lead to a black hole ``image'', and the uncertainties inherent in these observations themselves.
Here, we focus on the physics of the source of the measured light: the plasma surrounding the black hole.
Clearly, any test of GR using black hole images must be able to disentangle the physics of the emitting plasma from the effects of the underlying geometry, so it is important to highlight the difficulties and unknowns associated with the plasma physics.
We also briefly mention a few image simulation techniques and provide some possible starting points for the reader interested in exploring them further.

\bigskip

The photons observed by the EHT are assumed to be mainly emitted via synchrotron radiation by relativistic electrons in the accretion disk (or potentially from the jet region) \cite{EHT2019V,EHT2022V,Davelaar2023}.
The heating mechanism of the electrons in the accretion disk, and hence their temperature (which is essential for modeling their emission), is currently unknown.
It is unlikely that a single mechanism is at play everywhere in the accretion disk, and understanding the interplay between the most often conjectured heating mechanisms (such as turbulence, magnetic reconnection, and shocks) requires high-accuracy numerical simulations resolving all key physical processes. 

To calculate the temperature and dynamics of radiating electrons, a fully kinetic solution is required; that is, one would have to numerically solve the Vlasov equation\footnote{This is the Boltzmann equation for charged particles, where the general force term is replaced by the Lorentz force obtained from Maxwell's equations---see, e.g., \cite{Parfrey2019,Crinquand2022,Galishnikova2023a} for general relativistic kinetic simulations of black holes and \cite{Bacchini2019,Bacchini2020} for an explanation of the numerical methods.} in the curved spacetime of the black hole. However, a crucial difficulty in solving the Vlasov equation for the emitting electrons in the accretion disk plasma is that there is a large separation of scales between the size of the black hole and the typical plasma length scales, which are the Larmor radius (the radius of an electron's gyrating motion in a magnetic field) and the skin depth (the depth to which electromagnetic radiation can penetrate the plasma).
For example, for Sgr~A$^*$ (resp. M87$^*$), the gravitational radius is $r_{\rm g}=GM/c^2\simeq6.1\times10^{11}\,\rm{cm}$ (resp. $r_{\rm g}\simeq9.1\times10^{14}\,\rm{cm}$); whereas the nominal Larmor radius $r_{\rm L}\sim m_e c^2/\pa{|e|B_0}\sim\mathcal{O}(100\,\rm{cm})$,\footnote{This is for a typical magnetic field strength of $B_0\sim10\,\rm{G}$ in the accretion disk near the event horizon \cite{EHT2019V,EHT2022V}.} and the skin depth $d_{\rm e}=cm_{\rm e}^{1/2}/\pa{4\pi n_{\rm e}e^2}^{1/2}$ is of similar order as the Larmor radius $\mathcal{O}(500\,\rm{cm})$ for an electron number density $n_{\rm e}\simeq 10^6\,\rm{cm}^{-3}$ for Sgr~A$^*$ \cite{EHT2022V}, and $\mathcal{O}(3000\,\rm{cm})$ with $n_{\rm e}\simeq3\times 10^4\,\rm{cm}^{-3}$ for M87$^*$ \cite{EHT2019V}.
Kinetic simulations cannot resolve the (realistic) separation between these microscopic plasma scales and macroscopic scales at currently feasible numerical resolutions.
However, they can achieve the right hierarchy of scales ($r_{\rm g}\gg d_{\rm e}\gg r_{\rm L}$) for magnetized plasmas  \cite{Levinson2018,Parfrey2019,Crinquand2020,Chen2020,Crinquand2022,Galishnikova2023a}.

The most commonly used alternative to solving the Vlasov equation is to avoid having to resolve the microscopic plasma scales altogether by averaging over the particle distribution function.
This approach leads to a set of fluid equations enforcing the conservation of mass, momentum and energy of the plasma.
Coupled to the Maxwell equations in curved spacetime,\footnote{The spacetime is typically assumed to be stationary around supermassive black holes, which is reasonable since the backreaction of the accretion disk mass is negligible compared to the black hole mass, but in principle the spacetime metric could also be evolved by coupling the GRMHD equations to Einstein's equations; see, e.g., \cite{Duez2005,Shibata2005,Anderson2008,Mosta2014,Most2019,Cheong2020,Combi2023,Cook2023}.} the resulting set of evolution equations is called general-relativistic magnetohydrodynamics (GRMHD).

For typical EHT sources (i.e., low-luminosity active galactic nuclei), the plasma in the accretion disk and the black hole jet is collisionless, meaning that the mean free path to Coulomb collisions between charged particles is larger than the system size (set by the size of the accretion disk); see \cite{EHT2019V} for M87$^*$ and \cite{EHT2022V} Sgr~A$^*$.
The collisionless nature of the plasma also typically allows particles to accelerate and form a non-thermal energy tail, so that a Maxwell-Boltzmann distribution function does not accurately capture the plasma temperature. 
Therefore, electrons and ions in the plasma are not typically in thermal equilibrium and one cannot \emph{a priori} infer the electron temperature directly from the ion temperature.
However, the standard (ideal) form of the GRMHD equations assumes collisional plasma in thermal equilibrium; this describes the temperature and dynamics of the (heavier) ions, but results in the main uncertainty in calculating the essential electron temperature to describe the synchrotron radiation.

\bigskip

Despite their inability to capture the kinetic and collisionless plasma effects, GRMHD simulations do capture the global dynamics of the accretion flow and jet launching reasonably well (see, e.g., \cite{Porth2019} for a comparison of the most frequently used GRMHD codes). 
Even in ideal GRMHD, it is computationally challenging to resolve the scales on which energy dissipates and the plasma is heated, which is needed to probe sites where the observed radiation could be powered.
Typical GRMHD simulations employ numerical resolutions at which the magnetorotational instability (MRI) is at least minimally resolved \cite{Porth2019}---this is assumed to be one of the main drivers of the turbulence in the accretion disk in the standard turbulent accretion scenario, often called ``SANE'' (for Standard and Normal Evolution). 
In GRMHD simulations with 1,000 times higher resolution (10 times per direction) than typical EHT simulations, \cite{Ripperda2022} showed that features like magnetic reconnection and specific mechanisms driving the turbulence in the accretion disk can be resolved and result in observable radiation \cite{Hakobyan2023,Zhdankin2023}, while the overall dynamics remains similar to standard lower-resolution simulations.

The source of the plasma in the jet near its launching point just outside the event horizon is also not well understood.
GRMHD simulations have produced such jets and can predict their electromagnetic power \cite{Porth2019}.
Above a certain radius, in the open-field region of the jet, matter can escape the gravitational pull of the black hole and move outwards, whereas below that surface (the ``stagnation surface''), the plasma moves towards the event horizon.
This creates a vacuum region around the black hole.
In low-luminosity sources like M87, electron-positron pair production cascades will maintain a certain density (see, e.g., \cite{Levinson2018,Crinquand2020,Chen2020}).
However, in GRMHD the jet is magnetically insulated from the accretion flow, and as a result matter has to be artificially injected to maintain numerical stability.
Therefore, in the vicinity of the event horizon, GRMHD fails completely and a kinetic description is required to model the electron density, temperature, and emission.

Recently, \cite{Galishnikova2023a} was for the first time able to conduct a general-relativistic kinetic simulation of a simplified (yet relevant for Sgr~A$^*$; see, e.g., \cite{Ressler2020}) two-dimensional spherically infalling (accreting) plasma, and to compare it to a GRMHD simulation; it was found that there are many similarities in the global flow dynamics despite the differences caused by significant departures from thermal equilibrium (e.g., pressure anisotropies and heat fluxes not captured in GRMHD).

Some kinetic effects can be captured in GRMHD simulations by introducing macroscopic quantities (such as resistivity, viscosity, heat conduction, pressure anisotropies, etc.) or by evolving an additional equation for the electron temperature \cite{Bugli2014,Ressler2015,Chandra2015,Foucart2016,Chael2017,Chandra2017,Foucart2017,Qian2018,Ripperda2019,Vourellis2019,Ripperda2020,Most2021,Nathanail2022,Most2022,Vos2023}.
However, such subgrid models require careful benchmarking with kinetic simulations to test their validity (e.g., \cite{Bransgrove2021}).
These models typically require additional terms in the GRMHD equations describing the dissipation of energy and magnetic flux occurring on small scales, once again requiring high numerical resolutions to simulate these effects. 

\bigskip

In addition to major uncertainties due to collisionless and (underresolved) non-ideal plasma effects, the large-scale plasma flow geometry itself could also vastly differ from what is typically assumed and simulated---it is still unclear what the driver of the accretion is in the centers of galaxies.
In the Magnetically Arrested Disk models (``MAD'' \cite{Igumenshchev2003,Narayan2003}, as opposed to the SANE models mentioned above) favored by the EHT observations for both M87$^*$ and Sgr~A$^*$, a large ordered magnetic field forms near the event horizon.
In this MAD regime, the accretion flow goes through a relatively lengthy turbulent phase, increasing magnetic flux on the event horizon until the flux gets expelled quickly in a large eruption, after which the magnetic field forms a barrier that transiently halts the accretion \cite{BisnovatyiKogan1974,BisnovatyiKogan1976,Narayan2003}.
These flux eruptions are thought to play an essential role in the dynamics and geometry of the accretion flow \cite{Chatterjee2022}.
The flux eruptions may also have distinctive signatures in the EHT image \cite{Jia2023,NajafiZiyazi2023} and the photon ring \cite{Gelles2022}; counter-intuitively, they may even dim the image if the inner region emits predominantly at higher energies during an eruption.

All of the simulations mentioned above are initialized using artificial initial conditions, assuming a torus of weakly magnetized matter that becomes unstable and accretes onto the black hole.
In particular, the poloidal ($r,\theta$) magnetic field geometry and strength is tuned to obtain either the MAD or SANE scenario described above, whereas more realistic initial conditions are unknown and under-explored. 

It is for example unclear whether the strong field that is required to reach the MAD regime and launch a jet grows through a dynamo \cite{Liska2020,JacqueminIde2023} or is provided by the larger-scale environment.
For example, the accretion flow onto Sgr~A$^*$ is thought to be fed by stellar winds from the galactic center.
Depending on the properties of the winds, the inner accretion region may become MAD \cite{Ressler2020}, but the flow dynamics would differ significantly from the standard MAD simulations, which can have a distinguishable effect on the EHT image and variability \cite{Jia2022,Murchikova2022}.
For supermassive black holes in elliptical galaxies, like M87$^*$, pioneering work by \cite{Guo2023} showed that the inner accretion flow can significantly differ from the typically assumed MAD and SANE scenarios.
One interesting outcome from wind-fed accretion is the possibility of a tilted accretion disk with respect to the black hole spin axis \cite{Ressler2023}.
The effects of tilted accretion disks on the dynamics of the accretion flow have been explored extensively \cite{Chatterjee2020,White2020,White2022,Chatterjee2023} and can also result in distinguishable features in the images of M87$^*$ and Sgr~A$^*$.
Recently, another model was proposed by \cite{Blandford2022} exploring the possibility that the observed radiation would be produced in dissipative current sheets in an ergomagnetosphere, contrary to the standard MAD or SANE picture.
It is clear that the initial conditions used in standard simulations are artificial, and more astrophysically motivated accretion disk models have to be explored with GRMHD simulations to study the dynamics of the flow that leads to the observed emission.

\bigskip

To produce an image, the photon emission from the plasma must be simulated.
Typically, the distribution function of the emitting electrons (and hence their temperature) is put in by hand in a post-processing procedure.
Based on density, pressure, velocity and magnetic fields from a GRMHD simulation, emission and absorption properties are assigned to the radiating electrons, and a resulting Boltzmann equation for the emitted photons can then be solved.
A total intensity image is then produced by numerically ray tracing the photon trajectories in curved spacetime, solving along each ray the radiative transport equations that describe how the photons are emitted, absorbed, and scattered within the plasma.
The numerical methods for solving these equations are well-established and well-tested and can in principle be integrated to arbitrary accuracy (see \cite{Bacchini2018,Gold2020,Prather2023} for comprehensive comparisons between ray tracing methods and radiative transfer codes).
The main uncertainty in the simulated radiation and resulting image lies in the assumptions used for the electron temperature and its dependence on the GRMHD variables.\footnote{See two recently explored examples of kinetic effects on the EHT image from imbalanced (i.e., directionally dependent) turbulence preferentially heating ions \cite{Wong2023} and anisotropic electron distribution functions \cite{Galishnikova2023b}.}

\bigskip

GRMHD simulations are computationally expensive, typically requiring on the order of 1 million CPU hours per simulation (depending on resolution and integration time).
To find clear effects of the underlying geometry, one would in principle need to simulate many different GRMHD scenarios and identify universal features of the resulting images that are signatures of the geometry and not dependent upon the details of the plasma; see also \cite{Gralla2021,Lara2021,Ozel2022,Bauer2022} for more discussion of how to separate plasma effects from geometry.
For this reason, it can make sense to shortcut the GRMHD simulation step and instead use emission models that mimic to some extent the effects of GRMHD models, but instead are orders of magnitude less computationally expensive.
This can enable the quick simulation of many different emission scenarios and help identify such universal signatures of the underlying geometry.

One example of an emission-mimicking model is that of an equatorial and optically thin emission disk introduced in \cite{GLM2020} (see also \cite{CardenasAvendano2022} for a detailed description).
Geodesics that pass through the equator ``pick up'' extra intensity (representing the emission from the disk), resulting in a total brightness $I$ of a geodesic given by:\footnote{Polarization of the emission can also be simulated; see, e.g., Section~III.B in \cite{CardenasAvendano2022} for details.}
\begin{align}
    \label{eq:MockIntensity}
    I_{\rm o}=\sum_{n=0}^N\xi_n\,g^3\!\pa{r^{(n)}}\,I_{\rm s}\!\pa{r^{(n)},\phi^{(n)},t^{(n)}}\,,
\end{align}
where $I_{\rm s}$ is the local source intensity at the position where the light ray crosses the equatorial disk, $(r^{(n)},\phi^{(n)})$ is the location of the $(n+1)^\text{th}$ equatorial crossing, which occurs at time $t^{(n)}$, $g$ is the observed redshift (i.e., the relative redshift between the observer and the emitting plasma), $N+1$ is the total amount of times the geodesic crosses the equatorial plane, and $0<\xi_n$ is a ``geometric fudge factor'' that can be adjusted to mimic the effects of absorption or geometrical thickness of the emitting disk \cite{GLM2020,Chael2021,Paugnat2022,CardenasAvendano2022}.

A good choice for the local intensity function $I_s$ is such that its image \eqref{eq:MockIntensity} reproduces the main features seen in time-averaged GRMHD images.
Such a function, derived from Johnson's SU distribution, was introduced in \cite{GLM2020,CardenasAvendano2022}:
\begin{align}
    I_{\rm s}=\frac{e^{-\frac{1}{2}\br{\gamma+\sinh^{-1}\pa{\frac{r-\mu}{\sigma}}}^2}}{\sqrt{(r-\mu)^2+\sigma^2}}\,.
\end{align}
The tuneable parameters $\gamma$, $\mu$, and $\sigma$ can be used to simulate different types of emission \cite{Paugnat2022}.
The redshift factor $g$ is determined by a model for the fluid flow in the equatorial disk, which contains further tuneable parameters; see, e.g., \cite{CardenasAvendano2022} (especially Appendix B therein) for further details.\footnote{Temporal fluctuations can also be modeled using time-dependent emission functions in \eqref{eq:MockIntensity}; see for example \cite{Lee2021,CardenasAvendano2022}.}

\bigskip

Finally, keeping all of the above uncertainties and caveats in mind, we mention a few pedagogical starting points for the reader interested in simulating EHT images. 

A commonly employed open-source GRMHD code that the authors are familiar with is the BlackHoleAccretionCode ({\tt BHAC}) \cite{Porth2017,Olivares2019} (see \cite{Porth2019} for a comprehensive overview of GRMHD codes).
{\tt BHAC} solves the GRMHD equations in arbitrary spacetimes and coordinates, and has been extended to incorporate a non-ideal resistive GRMHD module \cite{Ripperda2019}. 
{\tt BHAC} has the ability to employ adaptive mesh refinement techniques with an oct-tree block-based approach, to increase the grid resolution in particular regions.
The code uses a second-order finite volume method and a constrained transport algorithm to maintain a divergence-free magnetic field \cite{Olivares2019}.
Typical simulations employ spherical Kerr-Schild coordinates (with $r$, $\theta$, and $\phi$ the radial, poloidal and toroidal angular coordinates, respectively\footnote{See, e.g., \cite{Davelaar2023} for a comparison between spherical and Cartesian Kerr-Schild coordinates.}) and use a torus in hydrodynamic equilibrium \cite{Fishbone1976} as initial condition.
A public release version, including documentation and example setups of MAD and SANE accretion disk scenarios, can be obtained from \url{https://bhac.science}.

We also mention two ray tracing codes that the authors are particularly familiar with and which are completely open-source.
First of all, the Adaptive Analytic Ray Tracing code \texttt{AART} \cite{CardenasAvendano2022} analytically integrates null geodesics in the Kerr geometry and can simulate both the stationary emission model described above around \eqref{eq:MockIntensity}, as well as time-dependent images.
The analytic integration method, as well as a nonuniform adaptive grid tailored to the lensing bands of Kerr, make it a very fast and efficient code for simulating properties of photon rings in Kerr (for lensing bands beyond Kerr, see \cite{CardenasAvendano2024}).
Another ray tracer that simulates stationary emission (but currently does not support time-dependent emission) is the Flexible Object-Oriented Ray Tracer \texttt{FOORT} \cite{FOORT}, which numerically integrates null rays in arbitrary geometries. \texttt{FOORT} is especially developed with maximal flexibility in mind, so as to be able to accommodate the study of other metrics, and allows for an easy simulation of images of beyond-GR objects.
Both \texttt{AART} and \texttt{FOORT} can further process images to produce simulated visibility amplitudes.
They are available at \url{https://github.com/iAART/aart} and \url{https://github.com/drmayerson/FOORT}.

\section{Brief overview of possible observables}
\label{sec:Observables}

In this Section, we will review a number of features of black hole images.
We will especially emphasize the extent to which these features have been argued to be \emph{universal} and \emph{robust} in the sense we discussed above; that is, to what extent each feature is controlled by the underlying geometry and whether or not it critically depends on the (unknown) details of the plasma emission.

\subsection{Black hole shadow and critical curve}
\label{sec:Shadow}

In the literature on black hole imaging (especially that which is focused on extracting geometrical data from such observations, or testing GR), the term ``shadow'' is used both ubiquitously and ambiguously.
Here, we attempt to give an overview of the possible definitions of the shadow, and its possible observable properties.
A pioneering work in the emission-dependence of the ``shadow'' is by Gralla, Holz, and Wald \cite{Gralla2019} (although we deviate somewhat from their original terminology here).

As discussed in Section~\ref{sec:Theory}, geodesics traced backwards from the observer screen towards the black hole can be divided into two categories: those that eventually fall into the black hole and those that escape to infinity in the far past.
The boundary separating those regions (which itself is strictly not part of either region) is the \emph{critical curve} and corresponds to asymptotically bound photon orbits around the black hole.
As we reviewed, the critical curve is well-known for Kerr \cite{Gralla2020,GrallaLupsasca2020c}.

A \textbf{``mathematical shadow''} could then be defined as the collection of points on the observer screen that lie inside the critical curve, i.e., those that correspond to geodesics that eventually fall into the black hole when traced backwards far enough.
Note that, by definition, the critical curve is an infinitely thin mathematical curve, and is \emph{not} in itself directly observable: depending on the astrophysical scenario, the emission may or may not produce a brightness deficit that coincides with it.

By contrast, a more physical notion of \textbf{``image shadow''} can be defined as the (generically present) central brightness depression in a black hole image.
More precisely, one could specify the ``outer shadow'' as the region interior to the bright ring appearing in black hole images (a separate feature known as the ``inner shadow'' \cite{Chael2021} will be discussed below).
This remains a rather ambiguous definition---for example: where do we consider the bright ring to ``end'' precisely?
In other words, how do we specify the boundary of the outer shadow?
This ambiguity only worsens when considering observations with less-than-perfect resolution (as in realistic situations) in which the surrounding (photon) ring is rather blurred.

Clearly, the mathematical shadow (as defined above) is a precise notion, whereas the image shadow is not.
It is important to note that the relation between these two notions of shadow depends on the details of the emission.
For example, emission coming from ``in front'' of the black hole will spoil the central brightness depression.
Only for very fine-tuned emission models will these two notions of shadow coincide: this can occur, for example, if the accretion flow around the black hole is spherically symmetric and optically thin \cite{Falcke2000,Narayan2019}.
However, in most cases, including those favored by EHT observations, these two notions of shadow will differ.
In principle, the discrepancy can be quite large \cite{Gralla2019}, though it has been argued that a large class of realistic emission models give (only) a small---yet still measurable---amount of deviation \cite{Bronzwaer2021}.
In any case, a measurement of the image (outer) shadow does not necessarily imply a measurement of the mathematical shadow or of the critical curve.

There is thus a tension between the notions of critical curve, which defines the mathematical shadow but is a non-observable theoretical construct, and the image feature of the (outer) shadow, which is observable but ambiguous (depends on the details of the emission) and does not necessarily correspond to the critical curve.
The outer shadow---and its implicit approximate equivalence with the mathematical shadow---is used in many observational contexts, but one should be careful not to mix the two concepts.
For example, the EHT uses the diameter of this central brightness depression as a measure for the mass of the observed black hole \cite{EHT2019VI}.
The analysis of \cite{Bronzwaer2021} investigated the image shadow and its (approximate) robustness over many different emission models, providing a partial justification for such measurements using the image shadow \emph{for Kerr}.

\bigskip

Nevertheless, the critical curve does still play a central role in black hole images.
The curve \emph{itself} is not observable, but the $n^\text{th}$ photon subring will converge (exponentially fast in $n$) to the critical curve, which may be viewed as the $n\to\infty$ photon subring (see Section~\ref{sec:PhotonRing}).
This means that resolving the diameter of a sufficiently high-$n$ photon subring will give an accurate estimate for the diameter of the critical curve.
It seems that in general, the $n=2$ photon subring is already close enough to the critical curve to be virtually indistinguishable from it (at least by eye, though high-resolution space-VLBI observations may still see the difference \cite{Paugnat2022}).
In fact, even though the $n\ge2$ photon rings and the critical curve (the limiting $n\to\infty$ ring) may not exactly coincide, they must follow the same \emph{shape} \cite{GLM2020,CardenasAvendano2023}.
More precisely, their \emph{projected diameter}---a notion defined in the Appendix---must follow a specific functional form known as a \emph{circlipse} (see further discussion in Section~\ref{sec:PhotonRingShape}).
This suggests an alternative pathway (as opposed to the measurement of the image shadow) for using the critical curve to characterize deformations of Kerr: one may quantify deviations from the Kerr prediction for the photon ring shape.

\bigskip

\begin{figure}[ht]
    \centering
    \includegraphics[width=\textwidth]{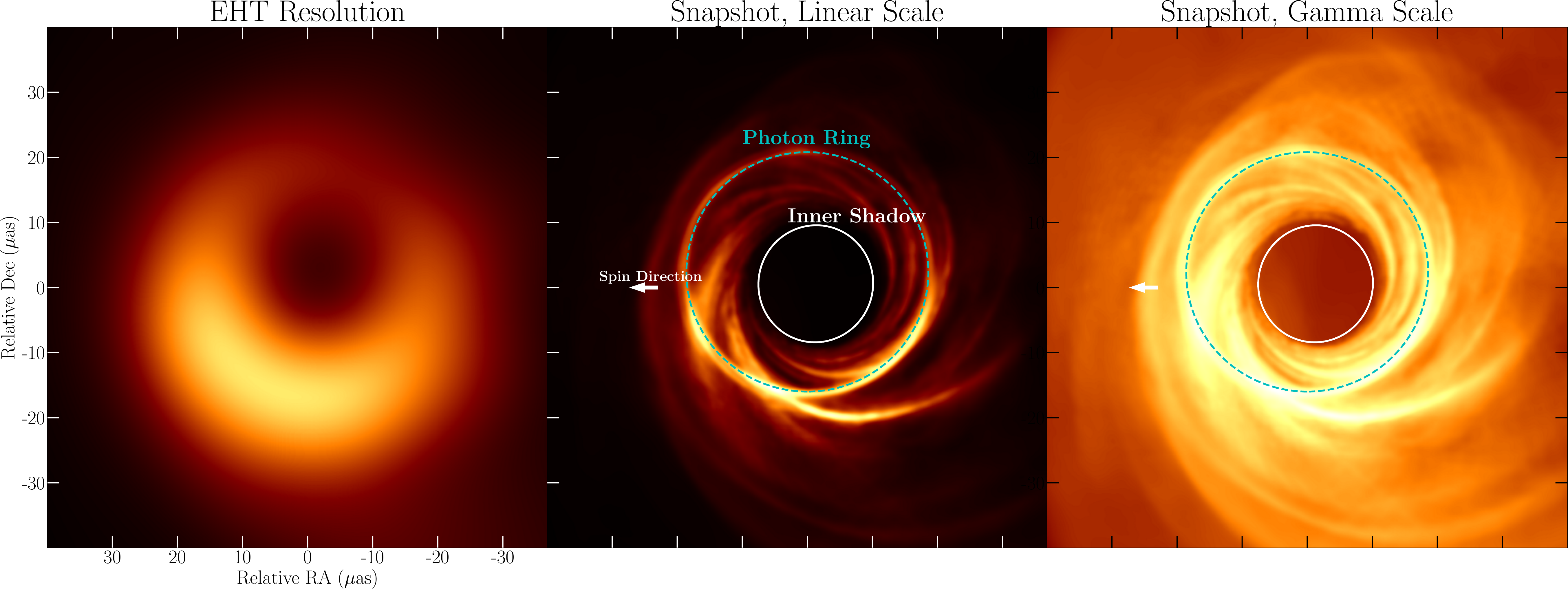}
    \caption{A simulated black hole image.
    \emph{Left:} Time-averaged image at EHT resolutions.
    \emph{Middle:} A GRMHD snapshot with the photon ring and inner shadow labeled.
    \emph{Right:} Same image in a gamma color scale that highlights the central brightness depression.
    This is Figure~1 of \cite{Chael2021}; further details about the image to be found therein.}
    \label{fig:InnerShadow}
\end{figure}

Finally, more recently the notion of (image) \textbf{``inner shadow''} was introduced \cite{Chael2021}.
It has a precise definition as the direct (weakly lensed) image on the observer screen of the equatorial event horizon.
This region, which is illustrated in Figure~\ref{fig:InnerShadow}, is markedly smaller than the critical curve or outer shadow.
As was argued in \cite{Chael2021}, the inner shadow seems to represent a region in the image that is consistently dark across emission models, and thus \emph{does} represent a universal signature in black hole images that can be measured and compared to the Kerr prediction.
However, the dynamic range (i.e., the observable range of brightness intensities) and resolution of the EHT is currently insufficient to resolve this inner shadow brightness depression; the necessary dynamic range could be achieved in the upcoming ngEHT \cite{Chael2021}.

\subsection{Photon ring interferometry}\label{sec:InterferometricPhotonRing}

As discussed above, the asymptotically bound photon orbits of Kerr trace out the critical curve on the viewer screen; this is a mathematical construct that is in principle unobservable.
By contrast, light rays that follow \emph{nearly} bound photon orbits can appear on the observer screen very close to the critical curve.
Their existence gives rise to an infinite sequence of photon subrings, with each successive ring produced by photons that executed an additional orbit around the black hole on their way from source to observer.
Assuming equatorial emission from an optically thin disk, the $n^\text{th}$ subring is formed by photons that crossed the equatorial plane $n$ times and thus collected $\sim n$ photons from the emission region (see also Section~\ref{sec:PhotonRing} above).

\begin{figure}[ht]\centering
\includegraphics[width=0.48\textwidth]{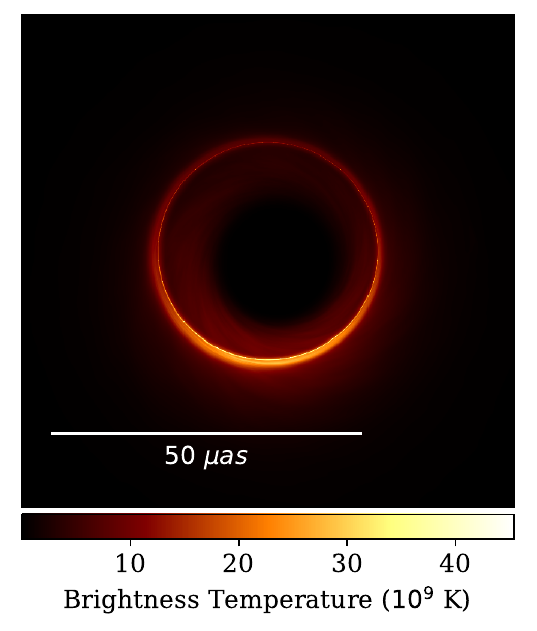}
\includegraphics[width=0.48\textwidth]{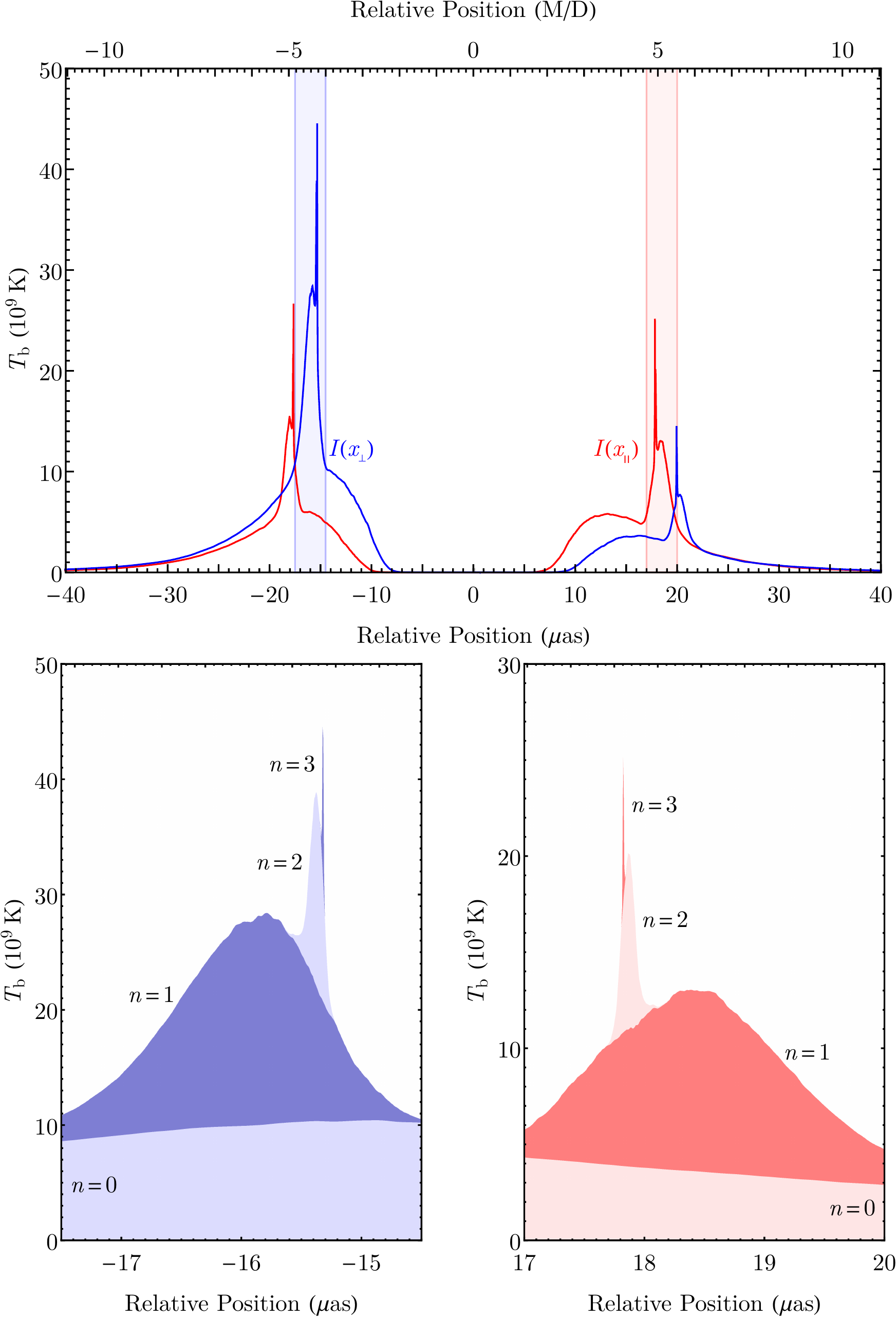}
    \caption{A simulated black hole image \emph{(left)}, and a cross-section of its intensity profile in which individual photon subring intensity peaks are clearly visible for $n=0,1,2,3$ \emph{(right)}.
    These are Figures~1 and 3 of \cite{JohnsonLupsasca2020}, reproduced under the terms of the CC Attribution 4.0 International License.}
    \label{fig:BrightnessProfiles}
\end{figure}

The photon rings are not only sharp features in the black hole image; they also produce distinctive interferometric signatures, which were first described in \cite{JohnsonLupsasca2020}.
Therein, it was argued that for ``large enough'' $n$, GR predicts that the subring flux ratio of successive subrings must be
\begin{align}
    \frac{F^{n+1}_{\text{ring}}}{F^{n}_{\text{ring}}}\approx e^{-\gamma}\,,
\end{align}
where $\gamma$ is the Lyapunov exponent (see Section~\ref{sec:PhotonRing}).
This statement essentially follows from the fact that successive images are demagnified by $e^{-\gamma}$, and it holds \emph{independently} of the details of the actual emission.
Figure~\ref{fig:BrightnessProfiles} illustrates how the photon ring decomposes into its characteristic subring structure, composed of a sequence of increasingly narrow and tall peaks (the successive photon rings).

An important insight of \cite{JohnsonLupsasca2020} is that this photon ring brightness profile gives rise to a universal behavior of the complex visibility $V(\mathbf{u})$ on long baselines.
In general, a ring with diameter $d$ and thickness $w\ll d$ contributes differently in two (asymptotic) regimes in the visibility \cite{JohnsonLupsasca2020}:
\begin{align}
    \textbf{(I):}\quad
    \frac{1}{d}\ll u\ll\frac{1}{w}\,,\qquad \textbf{(II):}\quad
    \frac{1}{d}\ll\frac{1}{w}\ll u\,.
\end{align}
Baselines in regime (I) resolve the diameter of the ring but not its thickness, so that the visibility in this regime behaves as if the ring is a (perfectly) thin ring, displaying damped periodic oscillations with weakly decaying envelope $|V(u)|\sim u^{-1/2}$ (see Section~\ref{sec:PhotonRingShape} below for the precise expected form of $V(\mathbf{u})$ in this regime).
On the (much) longer baselines of regime (II), both diameter and thickness are resolved, and the visibility is sensitive to the radial profile of the ring; in general, the visibility of a smooth ring in this regime (II) will decay exponentially, as the ring is ``resolved out'' \cite{JohnsonLupsasca2020}.
The width-to-diameter ratio for the $n=1$ subring is typically $w/d\lesssim 10\%$, and becomes exponentially smaller by $e^{-\gamma}\approx 5-10\%$ for higher-order subrings.

\begin{figure}[ht]
    \centering
    \includegraphics[width=\textwidth]{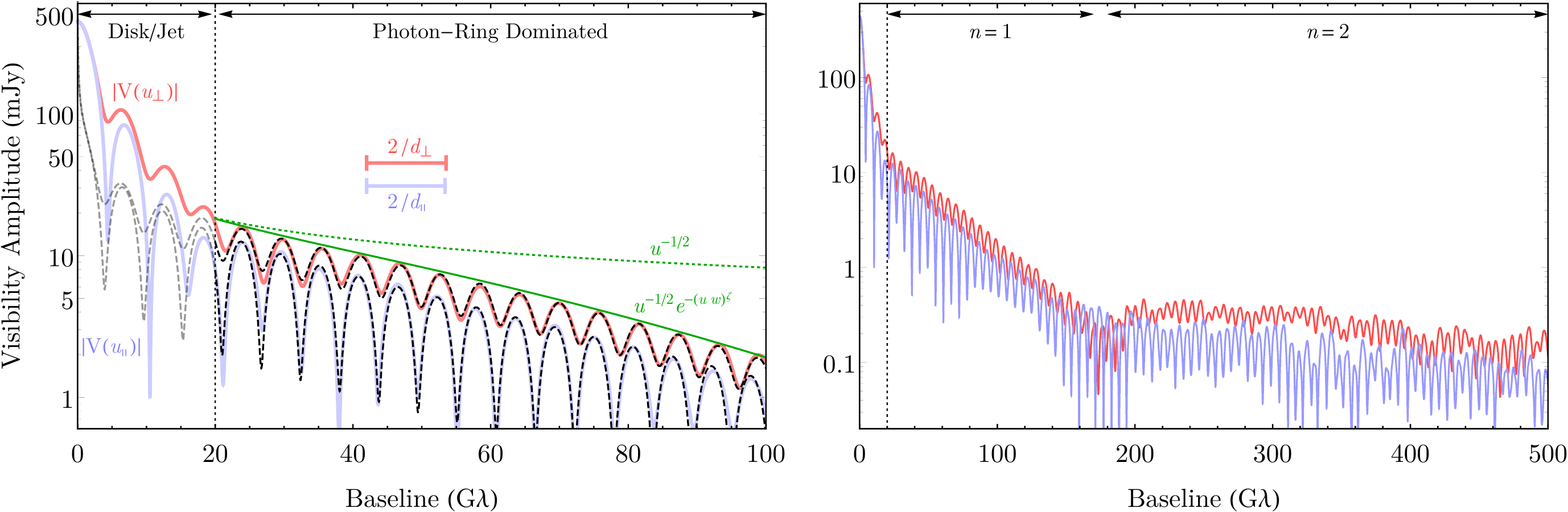}
    \caption{The visibility amplitude corresponding to the image in Figure~\ref{fig:BrightnessProfiles}, for baselines perpendicular (red) and parallel (blue) to the black hole spin axis.
    \emph{Left:} The universal photon-ring signature begins to dominate the visibility on long baselines.
    \emph{Right:} On even longer baselines, the signal from the $n=1$ is taken over by that of the $n=2$ subring.
    This is Figure~4 of \cite{JohnsonLupsasca2020}, reproduced under the terms of the CC Attribution 4.0 International License.}
    \label{fig:PhotonRingVisibility}
\end{figure}

The $n^\text{th}$ photon subring will then dominate the visibility signal in the regime:
\begin{align}
    \label{eq:SubringRegime}
    \frac{1}{w_{n-1}}\ll u \ll \frac{1}{w_n}.
\end{align}
Taken together, the subrings produce in the visibility a cascade of damped oscillations on progressively longer baselines, with each step in the cascade growing exponentially according to the Lyapunov exponent (since the widths of the subrings decrease as $w_{n+1}\approx e^{-\gamma n}w_1$).
Within each regime \eqref{eq:SubringRegime}, the visibility is dominated by a single subring and carries precise information about its diameter and thickness.
This is illustrated in Figure~\ref{fig:PhotonRingVisibility}, which displays a sample visibility amplitude along with its different subring-dominated baseline regimes.

Of course, the universal signatures of the photon subrings were stated above to hold for ``large enough'' $n$.
How big $n$ must precisely be to give a robust, emission-independent observable is unfortunately not entirely clear, and may depend somewhat on the precise photon ring feature being studied.
For example, for the photon ring shape discussed below in Section~\ref{sec:PhotonRingShape}, it has been comprehensively argued over a large range of emission models that $n=2$ is certainly ``large enough'' to extract universal data \cite{GLM2020}; even the $n=1$ subring seems to be relatively robust over different models \cite{CardenasAvendano2023}.

For a given photon ring observable, a careful analysis over many different emission models should be performed before concluding that a given $n^\text{th}$ subring gives ``universal'' results.
Existing analyses seem to suggest that $n\gtrsim 2$ and possibly even $n\sim 1$ should be high enough to extract interesting features from the photon ring.
This is at least promising from an observational perspective, since only the $n=1$ (and possibly $n=2$) subrings will potentially be resolved with near-future VLBI observations.

\subsection{Photon ring polarization}\label{sec:PhotonRingPolarization}

The preceding discussion in Section~\ref{sec:InterferometricPhotonRing} focused on the observed \emph{intensity} $I$, which constitutes only one of the electric field correlators that an interferometer measures (see Section~\ref{sec:Interferometry}).
The other cross-correlation functions of the electric field turn out to sample the Fourier transforms of the polarization Stokes parameters \cite{Roberts1994}, usually denoted $Q$, $U$ (the linear polarization components), and $V$ (the circular polarization).
Similarly to the intensity $I$, these polarimetric quantities display certain universal properties within the photon ring, which were first derived in \cite{Himwich2020}.

The linear polarization of light, which was introduced around \eqref{eq:PenroseWalker}, is encoded in the Stokes parameters $Q$ and $U$ as follows.
Define a two-dimensional vector $\vec{\mathcal{E}}$ on the screen with coordinates $(\alpha,\beta)$ of an observer at inclination $\theta_{\rm o}$ as:
\begin{align}
    \vec{\mathcal{E}}=\pa{\mathcal{E}_\alpha,\mathcal{E}_\beta}
    =\frac{\pa{\beta\kappa_2-\nu\kappa_1,\beta\kappa_1+\nu\kappa_2}}{\sqrt{\pa{\kappa_1^2+\kappa_2^2}\pa{\beta^2+\nu^2}}}\,,\qquad
    \nu=-\pa{\alpha+a\sin\theta_{\rm o}}\,,
\end{align}
where we used the complex Penrose-Walker constant $\kappa$ of \eqref{eq:PenroseWalker}.
Then the observed parameters $Q$ and $U$ combine into the complex polarization $P$ as
\begin{align}
    P=Q+iU=mIe^{2i\chi}\,,\qquad
    \chi=\tan^{-1}\pa{-\frac{\mathcal{E}_\alpha}{\mathcal{E}_\beta}}\,,
\end{align}
with $I$ the (Stokes parameter) intensity and $m$ the degree of polarization \cite{Himwich2020}.
The electric field vector $\vec{\mathcal{E}}$ encodes the direction of polarization of light; its sign is clearly unobservable and thus unphysical.
The polarization naturally decomposes into \cite{Himwich2020}
\begin{align}
    P=(\beta+i\nu)^2\mathcal{P}\,,\qquad 
    \mathcal{P}=\pa{\frac{mI}{\beta^2+\nu^2}}\frac{\overline\kappa}{\kappa}\,,
\end{align}
which leads to a remarkably simple universal pattern of the polarization within the photon ring.
The even subrings (resp. odd subrings) have equal linear polarization $\vec{\mathcal{E}}_{n+2}=\vec{\mathcal{E}}_n$, which implies that
\begin{align}
    \label{eq:EvenEvenPolarization}
    \mathcal{P}_{n+2}(d_{n+2},\varphi)=\mathcal{P}_n(d_n,\varphi)\,,
\end{align} 
where we expressed $\mathcal{P}=\mathcal{P}(\rho,\varphi)$ as a function of the polar coordinates $(\rho,\varphi)$ on the observer screen, and $d_n$ is the radius of the $n^\text{th}$ subring (at angle $\varphi$).
Moreover, the polarization gets complex-conjugated between odd and even subrings:
\begin{align}
    \label{eq:EvenOddPolarization}
    \mathcal{P}_{n+1}(d_{n+1},\varphi)=\overline{\mathcal{P}}_n(d_n,\varphi).
\end{align}

The striking universal polarization pattern of the photon ring is a stringent prediction of GR that can in principle be tested.
However, as in the discussion of Section~\ref{sec:InterferometricPhotonRing}, to accurately measure the $n^\text{th}$ subring's polarization, it is necessary to observe the source on very long baselines \eqref{eq:SubringRegime} dominated by the interferometric signature of the $n^\text{th}$ subring.
This implies that near-future observations will most likely only be able to measure the polarization of the $n=1$ (and possibly $n=2$) subrings, leaving tests of \eqref{eq:EvenEvenPolarization} beyond reach but perhaps opening the door to tests of \eqref{eq:EvenOddPolarization}.
Encouragingly, recent simulations \cite{Palumbo2022} suggest that \eqref{eq:EvenOddPolarization}, which was derived in the regime of ``large'' $n$, may even hold for the direct $n=0$ image and $n=1$ ring.
If so, then measuring this characteristic flip in the complex polarization could even lead to the first photon ring detection, before reaching the baselines needed to resolve it \cite{Palumbo2023}.

\subsection{Photon ring shape}
\label{sec:PhotonRingShape}

In Section~\ref{sec:InterferometricPhotonRing}, we described how the $n^\text{th}$ subring dominates the interferometric visibility within the baseline window \eqref{eq:SubringRegime}, namely over baseline lengths $u$ in the regime
\begin{align}
    \label{eq:SubringRegimeBis}
    \frac{1}{w_{n-1}}\ll u\ll\frac{1}{w_n}\,.
\end{align}
On such baselines, the $n^\text{th}$ ring can be approximated by a bright thin curve.
Using this insight, Gralla \cite{Gralla2020} showed that the visibility amplitude on long baselines \eqref{eq:SubringRegimeBis} takes the universal form
\begin{align}
    \label{eq:UniversalVisamp}
    \ab{V(u,\varphi)}\approx\sqrt{\frac{\pa{\alpha_\varphi^{\rm L}}^2+\pa{\alpha_\varphi^{\rm R}}^2+2\alpha_\varphi^{\rm L}\alpha_\varphi^{\rm R}\sin\pa{2\pi d_\varphi u}}{u}},
\end{align}
where $d_\varphi$ is the projected diameter of the curve, illustrated in Figure~\ref{fig:VisampSimulation} and precisely defined in the Appendix, while the coefficients $\alpha_\varphi^{\rm L,R}$ depend on the intensity profile around the curve.

Measuring the visibility amplitude on sufficiently large baselines and at different angles $\varphi$ in the baseline plane will enable the extraction of the projected diameter $d_\varphi$ as a function of polar angle $\phi_R=\varphi$ in the image---in other words, a measurement of the \emph{shape} of the $n^\text{th}$ photon ring.

\begin{figure}[ht]
    \centering
    \includegraphics[width=0.95\textwidth]{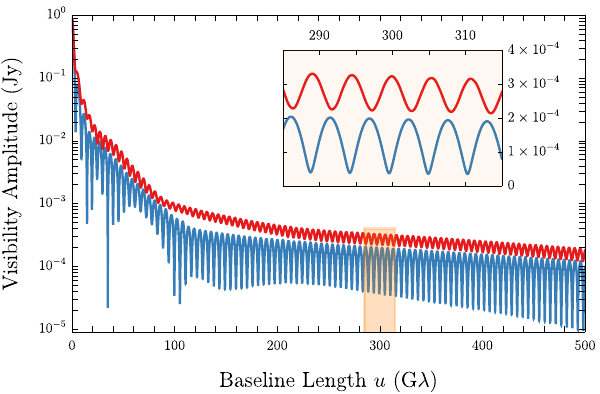}
    \caption{A simulation of the visibility amplitude at $\varphi=0^\circ$ (red) and $\varphi=90^\circ$ (blue) on long baselines corresponding to the black hole image in Figure~\ref{fig:CirclipseExample}.
    The inset displays the clear periodic ringing of \eqref{eq:UniversalVisamp} in the long-baseline regime \eqref{eq:SubringRegimeBis} where the $n=2$ subring dominates.
    This is part of Figure~4 from \cite{GLM2020}.}
    \label{fig:VisampSimulation}
\end{figure}

The shape of the \emph{critical curve} for the Kerr geometry was investigated in detail in \cite{GrallaLupsasca2020c}, where it was found that the projected diameter as a function of $\varphi$ is very well approximated by
\begin{align}
    \label{eq:Circlipse}
    \frac{d_\varphi^{\rm circlipse}}{2}=R_{\rm o}+\sqrt{R_1^2\cos^2{\varphi}+R_2^2\sin^2{\varphi}},
\end{align}
This is called the ``circlipse'' function as it is the sum of the diameter of a circle, $2R_{\rm o}$, with the diameter of an ellipse with axes $R_1$ and $R_2$.

\begin{figure}[ht]\centering
    \includegraphics[width=\textwidth]{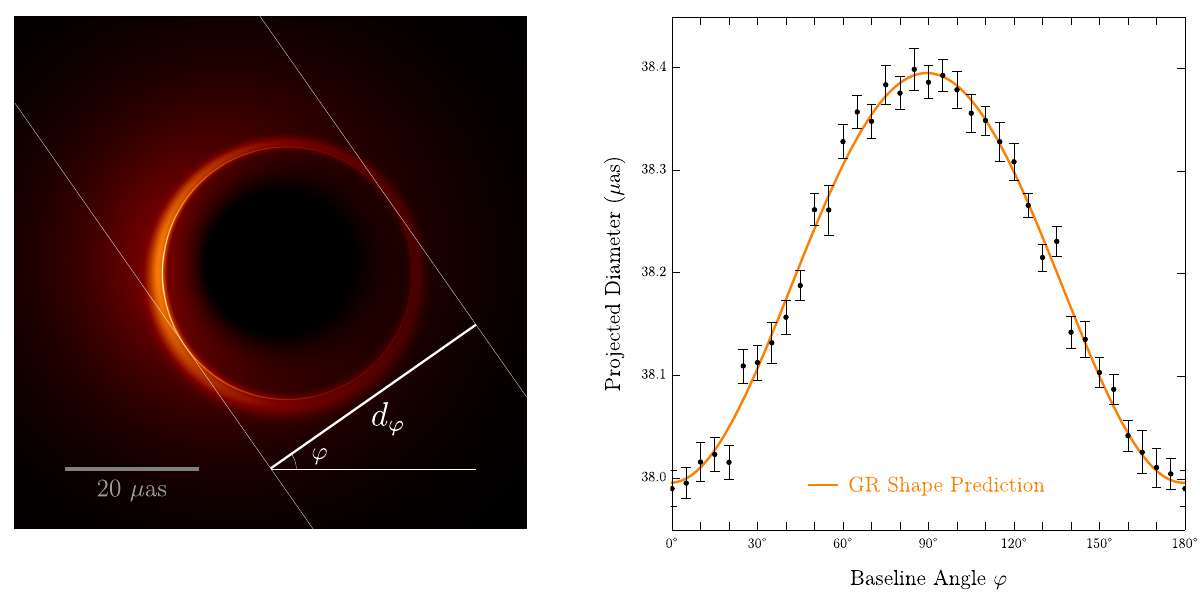}
    \caption{A simulated image of an equatorial disk around a Kerr black hole with spin $a/M=94\%$ viewed at inclination $\theta_{\rm o}=17^\circ$ \emph{(left)}, and the projected diameter extracted from the interferometric signature of $n=2$ photon ring at various values $\varphi$ around the baseline plane \emph{(right)}, with the best-fit circlipse shape \eqref{eq:Circlipse} shown in orange.
    This is Figure~1 from \cite{GLM2020}.}
    \label{fig:CirclipseExample}
\end{figure}

Since the photon subrings approach the critical curve at large $n$, it is reasonable to assume that for $n$ ``sufficiently'' large, the shape of the photon subrings will also be well-described by the circlipse.
This was the main argument of Gralla, Lupsasca, and Marrone \cite{GLM2020}, who showed that the $n=2$ subring shape (as extracted from mock interferometric data simulated from a wide range of models) can be expected to conform to the circlipse shape to a precision of $\lesssim0.04\%$ (see Figure~\ref{fig:CirclipseExample}).
This was confirmed by subsequent studies \cite{Paugnat2022,CardenasAvendano2022}.
More recently, it was shown that even a measurement of the $n=1$ subring shape would result in a circlipse, up to errors of order of a percent \cite{CardenasAvendano2023}.

In practice, it is technically challenging to extract the ring diameter $d_\varphi$ at different angles from (mock) interferometric data and find the best-fit circlipse function.
Such a fitting procedure was first developed in \cite{GLM2020} and later refined in \cite{Paugnat2022}, which provides a comprehensive and detailed explanation of an improved procedure and its complications.
To measure the $n=2$ ring and extract its projected diameter, it may be necessary to observe on baselines so large that a distant-Earth orbit mission would be required \cite{GLM2020}.
By contrast, the $n=1$ subring is already observable on relatively shorter baselines, and so is more likely to be observed in the near-future \cite{Gurvits2022,Kurczynski2022,CardenasAvendano2023}.

\subsection{Photon ring autocorrelations}\label{sec:PhotonRingAutocorrelations}

The light-emitting plasma around the black hole is expected to be a complex, turbulent medium, far from equilibrium.
As such, the emission that sources the photon rings will be subject to (brightness) fluctuations.
Different photons produced by the same fluctuation can travel along different paths around the black hole, circumnavigating it a different number of times, until they eventually end up in different photon subrings on the observer screen.
In other words, brightness fluctuations in the emission will produce correlated fluctuations across different subrings within the photon ring, reaching the observer at different times and appearing at different positions around the image \cite{Hadar2021,Wong2021}.

Given intensity fluctuations $\Delta I=I(t,\rho,\varphi)-\av{I(t,\rho,\varphi)}$, where $t$ is the observation time and $\rho,\varphi$ are polar coordinates on the observer screen, we can define the (integrated) \emph{two-point correlation function} (2PF) of intensity fluctuations as \cite{Hadar2021}:
\begin{align}
    \mathcal{C}(T,\varphi,\varphi')=\int\rho\ed\rho\int\rho'\ed\rho'\av{\Delta I(t,\rho,\varphi)\Delta I(t+T,\rho',\varphi')}\,,
\end{align}
where we assume that the fluctuations are statistically stationary, so that the correlator only depends on the temporal difference $T=t'-t$.
One can further integrate over angles to obtain the autocorrelation in time only (also known as the light curve autocorrelation), $\mathcal{C}_{\text{1D}}(T)=\int\ed\varphi\int\ed\varphi'\mathcal{C}(T,\varphi,\varphi')$.

\begin{figure}[t]
    \centering
    \includegraphics[width=\textwidth]{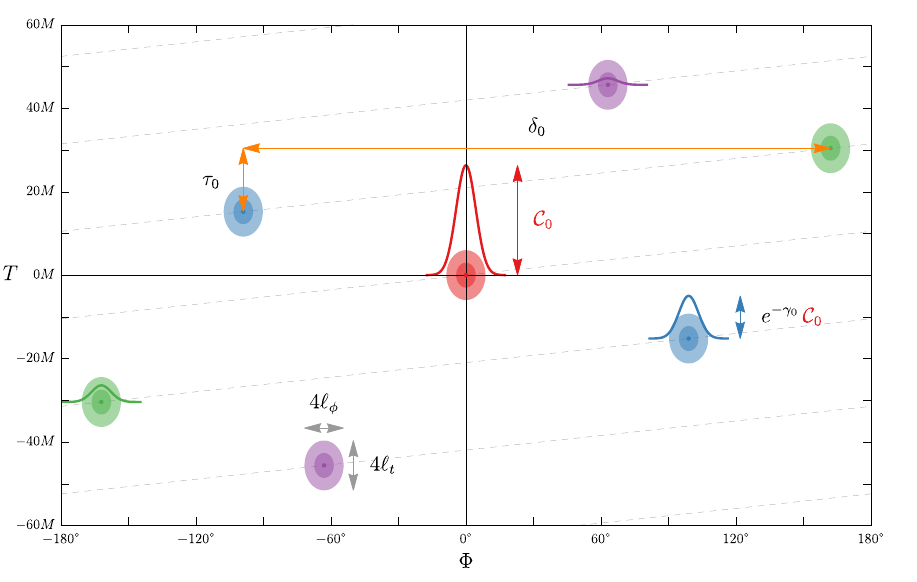}
    \caption{The characteristic structure of the photon ring autocorrelation function $\mathcal{C}(T,\varphi_0,\varphi_0+\Phi$) for a fiducial value $\varphi_0$.
    The relations between correlation peaks are determined by the critical exponents $\gamma$, $\delta$, and $\tau$ (shown here with a $0$ subscript denoting a polar observer) and therefore universal; on the other hand, the details $\mathcal{C}_0$, $l_t$, $l_\phi$ of the fluctuations are not universal as they are emission-dependent.
    This is Figure~1 from \cite{Hadar2021}; see the caption therein for additional details.}
    \label{fig:autocorrelations}
\end{figure}

The autocorrelation function $\mathcal{C}$ encodes the correlations between fluctuations on the ring that arise from photons emitted from approximately the same place (and time) but travelling on slightly different paths around the black hole before escaping to the observer.
As such, its behaviour is determined by the behaviour of the (nearly) bound photon orbits in the geometry and thus to a very good approximation by the critical exponents $\gamma$, $\delta$, and $\tau$ described in Section~\ref{sec:PhotonRing} \cite{Hadar2021}.
Indeed, the lensing behavior of a Kerr black hole produces a universal pattern of autocorrelation governed by these exponents, which is illustrated in Figure~\ref{fig:autocorrelations}.
These photon ring autocorrelations were first analyzed in \cite{Hadar2021} for certain emission models.\footnote{See \cite{Hadar2023} for a recent follow-up considering a more involved emission model, and also \cite{Chen2023} for an intriguing proposal for how deviations in this characteristic autocorrelation pattern could reveal new ultralight bosons.}

At least in principle, since the autocorrelation function is determined by the critical exponents, measuring these correlations effectively measures these exponents.
Since these exponents are extremely sensitive to changes of the underlying metric (see, e.g. ,\cite{JohnsonLupsasca2020,Staelens2023}), the autocorrelation function represents an exciting avenue for performing a precision test of the underlying black hole geometry.
It was estimated in \cite{Hadar2021} that, for M87$^*$ observations, detecting a strong signal in $\mathcal{C}$ may require many months or years of observations; with some optimism, a first estimate of $\mathcal{C}$ could be possible with EHT observations of M87$^*$ every few days over a span of a few months or years \cite{Hadar2021}. The situation may be (orders of magnitude) more optimistic for first estimates of $\mathcal{C}$ (i.e., the red and blue peaks in Figure~\ref{fig:autocorrelations}) for Sgr~A$^*$, although even there years of continuous observation would be necessary to resolve the second and higher-order correlation peaks (the purple and green peaks in Figure~\ref{fig:autocorrelations}) \cite{Hadar2021}.

\section{Testing general relativity in practice}\label{sec:Testing}

The previous Section focused on identifying possible observables in black hole images that could, in principle, give unambiguous information on the underlying black hole geometry.
Now, we turn to the question of how these observables could be leveraged into actual tests of GR.

To set the scene, we will first give a lightning overview of how and why GR can be expected to be corrected, and how possible deviations are modelled in practice.
Then, we will highlight the important difference between \emph{consistency} and \emph{discriminatory} tests of GR.
Finally, we give a few examples of both types of tests, and discuss the present and (near-)future prospects for such tests via VLBI observations.

\subsection{What could lie beyond general relativity?}\label{sec:BeyondEinstein}

The idea of testing GR naturally raises the question: what else could there be?
How could (or will) GR be corrected or changed?
Here, we review the main possibilities that are typically considered in beyond-GR phenomenology and especially in black hole imaging.

Black holes truly open a new window into the ``strong-field'' regime of gravity.
First of all, the Newtonian potential $GM/(Rc^2)$ becomes of order unity near the horizon.
Additionally, at least for relatively small (stellar-mass) black holes, the curvature scales involved near the horizon are much larger than any other curvatures that affect any measurements of gravitational observables.\footnote{Unfortunately, for the supermassive black holes relevant for black hole imaging, these curvature scales do become orders of magnitude smaller.}
See, e.g., \cite{Psaltis2019} for further discussion.

GR predicts its own downfall by requiring the existence of intrinsic curvature singularities inside of black holes, past which geodesics become nonextendable.
A slightly more subtle issue with black holes in GR is  the information paradox \cite{Mathur2009}, which arises from considering fluctuating quantum fields around a black hole \emph{horizon} (instead of at the singularity).

A natural assumption would be that a full, ultraviolet-complete theory of quantum gravity should somehow regularize the infinities at the black hole curvature singularities and resolve the information paradox satisfactorily.
Of course, it is notoriously difficult to quantize gravity, and the resolution of the various apparent paradoxes posed by black holes is certainly not a settled theoretical problem within GR---in particular, it is unclear what (observable) changes to make to classical GR.

Therefore, it is natural to instead characterize possible corrections to GR in an agnostic way.
An EFT approach---the gold standard in particle physics---would proceed by enumerating possible corrections to GR; in pure gravity, these are higher-derivative corrections in the action \cite{Endlich2017}:
\begin{align}
    S_{EFT}&=\frac{1}{16\pi G}\int\ed^4x\sqrt{-g}\left(R\right.\notag\\
    \label{eq:Corrections}&\left.+\lambda_{\rm ev}R\ind{_{\mu\nu}^{\rho\sigma}}R\ind{_{\rho\sigma}^{\delta\gamma}}R\ind{_{\delta\gamma}^{\mu\nu}}+ \lambda_{\rm odd}R\ind{_{\mu\nu}^{\rho\sigma}}R\ind{_{\rho\sigma}^{\delta\gamma}} \epsilon_{\delta\gamma\alpha\beta}  R\ind{^{\alpha\beta}^{\mu\nu}}\right),
\end{align}
where we have included the most general correction to GR involving at most six derivatives acting on the metric \cite{Endlich2017,Cano2019}.\footnote{This is true up to metric redefinitions of the form $g_{\mu\nu}\rightarrow g_{\mu\nu}+X_{\mu\nu}$ where $X_{\mu\nu}$ is a function of the curvature tensors.
Also, we have discounted total derivatives in the action as they do not contribute to the equations of motion.}
Four-derivative curvature corrections are total derivatives in four dimensions; however, if other matter fields such as a scalar are included as well, the four-derivative corrections can couple to these fields and become dynamical as well.
The most prominent examples of such four-derivative gravity theories coupled to scalars are Einstein-dilaton-Gauss-Bonnet (EdGB) theory and dynamical Chern-Simons (dCS) theory \cite{Cano2019,Alexander2009}.
By simulating what the observable effects of these higher-derivative couplings would be, bounds can then be put on the size of the couplings with current (and future) detections.
See, e.g., \cite{Okounkova2019} for a derivation of the critical curve in dCS theory, and \cite{Deich2023} for a recent study of the dCS deviations in the Lyapunov exponent.

\bigskip

The coupling constants $\lambda_{\text{odd,ev}}$ in \eqref{eq:Corrections} have dimension $[\text{length}]^4$ and so are naturally suppressed by powers of the Planck scale ($\sim 10^{-35}\,m$).
Higher corrections would come with even higher powers of the curvature and so would be even further suppressed.
The curvatures involved in observable properties of black holes are very far from the Planck scale: for a solar-mass black hole, the curvature scale at the (Schwarzschild) horizon is of the order of a kilometer; a supermassive black hole (such as M87$^*$ or Sgr~A$^*$) has curvatures at the horizon that are six or more orders of magnitude smaller.
The natural size of the coupling constants is then many orders of magnitude smaller than any reasonable bound that could ever be placed on them from any observations.
From an EFT perspective, therefore, the natural expectation is that observations will effectively never reveal deviations from GR due to these higher-derivative corrections.

However, this ``standard'' EFT reasoning may fail in a number of ways.
Of course, the ``natural'' scale for the higher-derivative corrections may not be the correct scale.
(Recall, for example, that the ``natural'' scale for the cosmological constant is about $10^{120}$ orders of magnitude larger than the actual observed value!)
We will mention three other important caveats to this EFT ``naturalness'' argument.
First of all, quantum effects may become important even at curvature scales well below where EFT predicts they should become appreciable.\footnote{Outside of the fuzzball paradigm, this idea has also recently gained a lot of traction in string theory cosmology, in the context of the ``Swampland'' program \cite{Palti2019}.}
In string theory, the \emph{fuzzball paradigm} \cite{Mathur2005,Mayerson2020,Mayerson2023} precisely hinges on arguments that the classical description of gravity breaks down at horizon scales---despite the absence of large curvatures.
Instead, at the horizon scale, quantum, stringy effects take over to render the black hole radically different than its classical GR description; instead of a featureless horizon, spacetime is replaced at these scales by stringy ``fuzz''; hence the ``fuzzball'' name.

A second caveat is that there may exist other dark matter fields that have so far been unobserved but nevertheless couple to gravity.
Effects of such dark matter would be especially relevant for the physics of (e.g.) black hole mergers \cite{Barausse2020}, where various interactions between the black holes and the dark matter can give rise to interesting signatures in the observed gravitational-wave signal.
In imaging phenomenology, the effects of (e.g.) dark matter halos around the observed supermassive black hole have been considered \cite{Hou2018}, as have the effects of ultralight bosons on the autocorrelation pattern of the photon ring \cite{Chen2023}.

A third caveat is that there may exist other, hitherto-unknown ultracompact exotic compact objects (ECOs).
Such ECOs could be more compact than a neutron star but still less compact than a black hole, and have evaded detection up until now due to their compactness and lack of emission.
This is also related to the first and second point above: a fuzzball is an example of a possible ECO, and ECOs often rely on the existence of additional (dark) matter fields to support their compact structure.
For an overview of possible ECOs and their phenomenology, see \cite{Cardoso2019}.

\bigskip

In practice, there are a number of ways to analyze and parametrize possible strong-gravity phenomena beyond GR.
A well-traveled path in black hole imaging is to parametrize deviations from the Kerr metric itself, without worrying about how such metric deviations represent a solution to any underlying theory.
By construction, such parametrizations are then model-agnostic.
An example of this is the Johannsen metric \cite{Johannsen2013,Johannsen2011}, which parametrizes the most general deviation from Kerr that still allows for separable geodesic equations.

Alternatively, one can consider solutions to alternative or corrected theories of gravity.
For example, one can consider all possible pure gravitational corrections to GR arising from \eqref{eq:Corrections}.
It is also possible to take a theory of GR coupled to (dark) matter and consider solutions of such theories that are similar but different to Kerr.
Perhaps the simplest example is the Kerr-Newman black hole, which is a solution to Einstein-Maxwell theory.
Another example is the theory obtained by minimally coupling a scalar field to GR, which leads to the existence of ``hairy'' black holes \cite{Herdeiro2014,Herdeiro2015} as well as horizonless ECOs called boson stars.
The fuzzballs mentioned above are also a type of ultracompact ECO \cite{Mathur2005,Mayerson2020}; these are complex solutions to certain supergravity theories (i.e., theories of gravity coupled to various scalar and gauge fields)\footnote{More precisely, the fuzzballs that can be represented by classical solutions to a supergravity theory are called microstate geometries or microstate solutions \cite{Bena2023}.} and are posited in string theory to be the individual microstates of a black hole---meaning that counting these microstates should reproduce the black hole's entropy.
Images of boson stars and hairy black holes have been studied in, e.g., \cite{Cunha2015,Cunha2016,Vincent2016,Olivares2020}; rudimentary fuzzball imaging has been analyzed in \cite{Bacchini2021}.

The above approaches all involve classical gravity theories or geometries, and are essentially devoid of true quantum corrections to gravity.
Investigating (measurable) quantum effects in imaging is also more or less unexplored territory, with perhaps a few notable exceptions such as \cite{Giddings2018}, although even there the quantum effects are posited to lead to observable coherent (classical) fluctuations of the metric.

\subsection{Consistency tests versus discriminatory tests}
\label{sec:NullTests}

Testing GR in observations goes hand-in-hand with looking for beyond-GR effects, either in the form of corrections to black hole metrics or new ECOs as discussed in the previous Section.
One can make a distinction between two different types of tests of GR:
\begin{itemize}
    \item \textbf{Consistency tests, or null tests}, in which GR is shown to be consistent (or not) with the observed data up to a certain precision level; and
    \item \textbf{Discriminatory tests}, in which GR is tested \emph{against} another, different theory, and it is shown that the observed data prefers either GR or the alternative theory.
\end{itemize}
In statistical terms, consistency tests correspond to a hypothesis test  where the null hypothesis is GR and the alternative hypothesis is not-GR, whereas a discriminatory test could be a Bayesian model selection between GR and the alternative theory.

For example, GR predicts that the shape of the photon ring is a circlipse (see Section~\ref{sec:PhotonRingShape}).
Testing if and how well a measured photon ring shape fits the circlipse functional shape is then a \emph{consistency} test of GR.
However, the circlipse shape of the photon ring is fairly universal across many models of corrections to black hole physics (see \cite{Staelens2023} for examples and \cite{universality} for an explanation why this is the case): in other words, Kerr \emph{and} many other black holes beyond GR will all have photon rings that have circlipse shapes.
This implies that fitting an observed photon ring shape to the circlipse functional form is not necessarily a good discriminatory test, since it has little distinguishing power between different theories.

\bigskip

Existing EHT observations are consistent with the theoretical modelling and expectations based on GRMHD-simulated accretion flows (see Section~\ref{sec:Plasma}) onto a Kerr black hole \cite{EHT2019I,EHT2019VI}.
In this sense, the EHT measurements are certainly a \emph{consistency} test of GR.
There has been vigorous debate as to the level of precision of this test \cite{Gralla2021} and how this level should even be defined.
We take the view that the test is both obviously valuable, and also obviously rather imprecise (see, e.g.. Sections \ref{sec:Interferometry}, \ref{sec:Plasma} and \ref{sec:Shadow}).
We are nearing a new era of precision black hole imaging (see below Section~\ref{sec:Future}) in which much greater precision will be achieved in observations and thus in the level of precision in the (consistency) testing of GR.

\bigskip

It is important to stress that the level of precision that is achievable with a given test (whether a consistency or discriminatory test) depends on both the measurement precision, as well as the inherent uncertainties associated with the test.
For example, we discussed the (image) shadow above (Section~\ref{sec:Shadow}).
The existence of such a shadow may be relatively robust over different plasma emission models \cite{Bronzwaer2021}, but the precise details---such as its size, for instance---do depend (however slightly) on the emission details.
As long as that is true, there will be an inherent uncertainty associated with any test based on the details of the image shadow.
Moreover, this uncertainty does not improve when better observations are made, except as a secondary effect of being able to better constrain the possible emission models.

\subsection{What is (not) a test, with examples}\label{sec:testexamples}

In this Section, we give some examples of proposed tests, both of consistency and discriminatory.
We also emphasize the most common pitfalls associated with such proposals.
Our aim is not to be comprehensive in giving an overview of all proposed imaging-based tests of GR, but rather to sketch of the current status of the field.

\bigskip

Given a black hole that admits separable null geodesics, an simple analysis to carry out is to study how its critical curve deviates from Kerr; see, e.g.,\cite{Johannsen2010,Bambi2010,Abdujabbarov2013,Amarilla2013,Vincent2016,Giddings2018,Mureika2017,Chen2020,Uniyal2023b,Uniyal2023b} for a non-exhaustive list of examples.
However, as in the above discussion in Section~\ref{sec:Shadow}, we want to stress that this is only a first indication of how such deformations of Kerr \emph{might} be observable, since the critical curve is not directly observable.
For example, the critical curve will typically be close to, but not precisely at, the boundary of the image shadow \emph{for Kerr}---but this is not obviously still the case in a deviating black hole metric.
To provide further evidence of observable consequences of a change in critical curve, a next step could be an analysis generalizing the Kerr study of \cite{Bronzwaer2021}, i.e., analyzing the image shadow  across different emission models on the alternative black hole metric in question.

The photon ring should also closely approximate the critical curve (although this should, once again, be verified for the alternative black hole).
However, as discussed in Section~\ref{sec:PhotonRingShape}, only the projected diameter is really relevant for the measurement of the photon ring (and thus critical curve) shape, so even seemingly quite different critical curves may give rise to the same interferometrically observed shape.
In other words, to (only) look at the critical curve of a beyond-Kerr metric is certainly a good \emph{starting point} to understand deviations of beyond-Kerr metrics in black hole imaging, but this analysis does not yet result in a good understanding of what interesting signals could arise from such a beyond-Kerr metric in interferometric observations.\footnote{Essentially the same reasoning applies to any rudimentary imaging simulations such as a four-color screen (e.g., \cite{Cunha2016,Bacchini2021}).}

The above example also illustrates a subtle pitfall in black hole imaging observations: a good test of GR should not revolve around a property of the \emph{image} itself---rather, it should follow from a property of the \emph{visibility} (i.e., the Fourier transform of the image), or even the visibility amplitude, since that is what is actually observed (see Section~\ref{sec:Interferometry}).
Our intuition is not always well-suited to understanding how image features translate to features in the Fourier-transformed visibility (or vice versa), and features that seem distinctive in an image may not be so easily distinguished in the visibility.

\bigskip

With these subtle pitfalls in mind, we can discuss a few examples of proposed consistency and discriminatory tests.
As aforementioned, the EHT observations provide a prime example of consistency tests.
In particular, the observed mass and spin parameters of both M87$^*$ and Sgr~A$^*$ are well within the expected range from other observations \cite{EHT2019I,EHT2022I}.
Measuring the inner shadow (see Section~\ref{sec:Shadow}) is a proposed consistency test that would be possible with the ngEHT \cite{Chael2021}.

Various consistency tests involving the photon rings have been proposed and studied in detail, including an estimate on (near-)future obtainable precision of these tests.
Examples include \cite{GLM2020} for the shape of the $n=2$ ring (see Section~\ref{sec:PhotonRingShape}), \cite{Paugnat2022} for further analysis of the $n=2$ ring shape over a range of simulated emission models, \cite{CardenasAvendano2023} for the observed shape of the $n=1$ ring.
The photon ring autocorrelations (see Section~\ref{sec:PhotonRingAutocorrelations}) were discussed as a consistency test in \cite{Hadar2021}.

\bigskip

Currently developed discriminatory tests are fewer in number.
There was a strong claim that measurements of the (image) shadow can be used to constrain post-Newtonian (PN) deviations to the metric \cite{Psaltis2020}, although it is not obvious how well-suited such a PN-expansion is for describing geodesics near bound orbits, as the PN expansion precisely breaks down near the horizon \cite{Volkel2021}; see also \cite{Gralla2019,Gralla2021}.
Other analyses show that constraints on beyond-Kerr parameters in the metric would only be rather weak \cite{Ayzenberg2022}, and that specifically the only \emph{type} of models that could realistically be constrained are those with dimensionless parameters (and so certainly not corrections such as in \eqref{eq:Corrections}) \cite{Glampedakis2021}.

GRMHD simulations are quite computationally expensive.
It is therefore perhaps remarkable that GRMHD plasma simulations have already been performed in a few models of beyond-GR compact objects, with the aim of investigating possible deviations in imaging observations.
The difference in images of a Kerr versus a dilatonic black hole was analyzed in \cite{Mizuno2018,Fromm2021,Roder2023}; a conclusion from these works is that it is again very difficult with current observation capabilities to distinguish Kerr black holes from other objects. 
In \cite{Olivares2020}, accretion flows onto horizonless boson stars were considered, leading to certain possible discriminatory signatures in imaging observations.
Spherical accretion in certain alternative gravity theories was analyzed in \cite{Bauer2022}.
Relatedly, GRMHD simulations in a large class of non-vacuum spacetimes was considered in \cite{Kocherlakota2023}.
Finally, the EHT collaboration has tested some basic image properties of a number of analytic metrics in the context of the first Sgr~A$^*$ observations \cite{EHT2022VI}.

It is perhaps somewhat surprising how few attempts have been made at discriminatory tests involving the photon rings, given our focus on observables involving them in Section~\ref{sec:Observables}.
The theoretical derivation of the photon rings in Kerr is crucially linked to the critical curve corresponding to bound photon orbits (see Section~\ref{sec:PhotonRing}).
For geometries beyond Kerr, the separability of the geodesic equations (generically) should not hold, rendering the entire reasoning that led to the derivation of the critical curve moot---does a single, connected critical curve exist for generic beyond-Kerr black holes (or other ultracompact objects)?
Probably not \cite{Staelens2023}.
What are the effects on the various properties of the photon rings?
And even for metrics allowing for separable geodesic equations, how are the properties of the photon ring altered, such as its shape and critical exponents?
A first attempt to answer such questions was made in \cite{Staelens2023} in the context of the Lyapunov exponent and photon ring shape for simple black hole metrics admitting separable geodesic equations.
As mentioned above, it was found that the photon ring circlipse shape, while a good consistency test of GR, is too generic to be able to distinguish between different models; the underlying reason for this universality was further exposed in \cite{universality}.
Another recent advance was the development of a formalism to define and calculate the Lyapunov critical exponent for generic metrics without geodesic separability \cite{Deich2023}.
Finally, very recently a discriminatory test was proposed based on the \emph{lensing band} structure of a beyond-Kerr spacetime: in essence, the idea is that if a photon ring is observed, its position can be used to rule out models where the observed photon ring does not fit in the model's lensing band \cite{CardenasAvendano2024}.
Evidently, there are still many photon ring properties that have not been considered yet in any discriminatory tests, leaving numerous interesting features of beyond-GR photon ring physics as yet to be explored.

\subsection{Where we are now and what the future holds}\label{sec:Future}

Current EHT observations show a bright ring of diameter around $\sim40-50\,\mu$as (for both M87$^*$ and Sgr~A$^*$) surrounding a dark brightness depression.
As we have discussed, this is already a type of consistency test of GR (see Sections~\ref{sec:NullTests} and \ref{sec:testexamples}).

The ngEHT will upgrade and expand the existing EHT telescope array with additional telescopes, improved observation methods, and greatly expanded observation durations \cite{Johnson2023}.
Key science goals for the ngEHT include measuring properties of the photon ring as well as measuring the inner shadow (as a signature of the existence of the event horizon; see Sections \ref{sec:Shadow}) \cite{Ayzenberg2023}.

The current observations resolve the $n=0$ image, using photon ring terminology.
However, the actual photon ring comprised of light that orbited around the black hole (i.e., the $n\geq 1$ images) has not yet been resolved,\footnote{There has been a claim of a detection of the photon ring in the EHT M87$^*$ observations \cite{Broderick2022}, although this claim is heavily disputed \cite{LockhartGralla2022a,LockhartGralla2022b,Tiede2022} (and in our opinion manifestly wrong).} and requires longer baselines than can be accessed with a solely Earth-based telescope array.
Future space-based VLBI missions \cite{Gurvits2022,Kurczynski2022} aim to extend the EHT to space via an orbiting satellite, so as to sample longer baselines and enable a measurement of the photon ring and some of the associated observables discussed in Sections \ref{sec:InterferometricPhotonRing}, \ref{sec:PhotonRingPolarization}, \ref{sec:PhotonRingShape}, and \ref{sec:PhotonRingAutocorrelations}.
Such missions will open the door to more precise tests of GR in the strong-field regime.
As we have discussed, the strong tests that are and will be possible are mostly \emph{consistency} tests of GR and the Kerr metric, although there is also some potential for \emph{discriminatory} tests using beyond-GR models.

The EHT has given us a first unprecedentedly close view of the horizon-scale emission from the close environment of two supermassive black holes.
It is clear that observing light originating from and affected by the strong-field gravitational pull of a black hole is a compelling and necessary experiment to carry out and improve.
Besides providing a way to qualitatively confirm predictions of strong-field GR, such experiments may also uncover crucial information about black hole spin, jet-launching, electromagnetic energy extraction, and other astrophysical phenomena of great interest.
The future of black hole imaging is bright.

\begin{acknowledgement}
We would like to thank F. Bacchini, A. C\'ardenas-Avenda\~no, and S. Gralla for interesting discussions.
A.L. is supported by NSF grants 2307888 and 2340457.
D.R.M. is supported by Odysseus grant G0H9318N of FWO Vlaanderen.
B.R. acknowledges support from the Natural Sciences \& Engineering Research Council of Canada (NSERC) and from the Simons Foundation (grant MP-SCMPS-00001470).
S.S. is supported by the Centre for Doctoral Training (CDT) at the University of Cambridge funded through STFC.
This work is also partially supported by the KU Leuven C1 grant ZKD1118 C16/16/005.
\end{acknowledgement}

\section*{Appendix: Projected diameter}
\label{app:ProjectedDiameter}
\addcontentsline{toc}{section}{Appendix: Projected diameter}

Here, following \cite{GrallaLupsasca2020c}, we give the appropriate definition of projected diameter that is used in Section~\ref{sec:PhotonRingShape}.

Given a closed curve in a plane, parametrized as $(x(\sigma),y(\sigma))$, one can define the normal angle to the curve by:
\begin{align}
    \tan{\varphi(\sigma)}=-\frac{x'(\sigma)}{y'(\sigma)}.
\end{align}
For convex curves, this can be inverted to obtain $\sigma(\varphi)$ and obtain a normal-angle parametrization $(x(\varphi),y(\varphi))$ (see \cite{GrallaLupsasca2020c} for a prescription for non-convex curves).
The \emph{projected position} of the curve is then defined as:
\begin{align}
    f(\varphi)=x(\varphi)\cos{\varphi}+y(\varphi)\sin{\varphi},\qquad
    \varphi\in[0,2\pi).
\end{align}
Interestingly, this function fully encodes the curve, which may be reconstructed via
\begin{align}
    x(\varphi)=f(\varphi)\cos{\varphi}-f'(\varphi)\sin{\varphi},\qquad
    y(\varphi)=f(\varphi)\sin{\varphi}+f'(\varphi)\cos{\varphi}.
\end{align}
We can further decompose the projected position  into its parity-even and parity-odd components:
\begin{align}
    d_\varphi=\frac{f(\varphi)+f(\varphi+\pi)}{2},\qquad
    C_\varphi=f(\varphi)-f(\varphi+\pi),\qquad
    \varphi\in[0,\pi),
\end{align}
where $d_\varphi$ is called the \emph{projected diameter} and $C_\varphi$ is the \emph{projected centroid} of the curve (see, e.g., Figure~3 in \cite{GrallaLupsasca2020c} for further illustration).

In an elegant exercise in Fourier analysis, Gralla \cite{Gralla2020} showed that a bright and narrow curve with projected diameter and centroid $(d_\varphi,C_\varphi)$ produces an interferometric response along the polar angle $\varphi$ on the observer screen given by:
\begin{align}
    V(u,\varphi)\approx\frac{e^{-2\pi iC_\varphi}}{\sqrt{u}}\pa{\alpha_\varphi^{\rm L}e^{-\frac{i\pi}{4}}e^{i\pi d_\varphi u}+\alpha_\varphi^{\rm R}e^{\frac{i\pi}{4}}e^{-i\pi d_\varphi u}},
\end{align}
The resulting visibility amplitude $|V(u,\varphi)|$, which is more robust to noise than the phase, is then given by \eqref{eq:UniversalVisamp}, which only depends on the projected diameter $d_\varphi$.

\bibliographystyle{utphys}
\bibliography{imagingchapter}

\end{document}